\documentclass[
  journal=pasa,
  manuscript=research-paper,
  year=2023,
  volume=37,
]{cup-journal}

\usepackage{graphicx}

\usepackage{amsmath}
\usepackage[nopatch]{microtype}
\usepackage{booktabs}

\usepackage{hyperref}

\definecolor{darkblue}{rgb}{0.0, 0.0, 0.5}
\definecolor{darkred}{rgb}{0.8, 0.0, 0.0}

\hypersetup{colorlinks=true,citecolor=darkblue,linkcolor=darkblue}

\makeatletter
\DeclareRobustCommand{\HI}{%
  \mbox{H\check@mathfonts\fontsize\sf@size\z@\selectfont I}%
}
\makeatother

\newcommand\arcsec{\mbox{$^{\prime\prime}$}}%

\def\farcs{%
 \mbox{%
  \kern  0.13ex.%
  \kern -0.95ex\arcsec%
  \kern -0.1ex%
 }%
}%

\newcommand{\rev}[1]{\textcolor{black}{#1}}

\title{The Galaxy Zoo Catalogs for Galaxy And Mass Assembly (GAMA) Survey}

\author{Benne W. Holwerda}
\affiliation{University of Louisville, Department of Physics and Astronomy, 102 Natural Science Building, 40292 KY Louisville, USA.}

\author{Clayton Robertson}
\affiliation{University of Louisville, Department of Physics and Astronomy, 102 Natural Science Building, 40292 KY Louisville, USA.}

\author{Kyle Cook}
\affiliation{University of Louisville, Department of Physics and Astronomy, 102 Natural Science Building, 40292 KY Louisville, USA.}

\author{Kevin A. Pimbblet}
\affiliation{E.A.Milne Centre for Astrophysics, University of Hull, Cottingham Road, Kingston-upon-Hull, HU6 7RX, UK}

\author{Sarah Casura}
\affiliation{Hamburger Sternwarte, Universit\"at Hamburg, Gojenbergsweg 112, 21029 Hamburg, Germany}

\author{Anne E. Sansom}
\affiliation{Jeremiah Horrocks Institute, University of Central Lancashire, Preston PR1 2HE, UK}

\author{Divya Patel}
\affiliation{University of Louisville, Department of Physics and Astronomy, 102 Natural Science Building, 40292 KY Louisville, USA.}

\author{Trevor Butrum}
\affiliation{University of Louisville, Department of Physics and Astronomy, 102 Natural Science Building, 40292 KY Louisville, USA.}

\author{David H. W. Glass}
\affiliation{Jeremiah Horrocks Institute, University of Central Lancashire, Preston PR1 2HE, UK}

\author{Lee Kelvin}
\affiliation{Department of Astrophysical Sciences, Princeton University, 4 Ivy Lane, Princeton, NJ 08544, USA}
\alsoaffiliation{Department of Physics, University of California, One Shields Ave., Davis, CA 95616, USA}
\alsoaffiliation{Astrophysics Research Institute, Liverpool John Moores University, IC2, LSP, 146 Brownlow Hill, Liverpool, L3 5RF, UK}

\author{Ivan K. Baldry}
\affiliation{Astrophysics Research Institute, Liverpool John Moores University, IC2, Liverpool Science Park, 146 Brownlow Hill, Liverpool, L3 5RF, UK}

\author{Roberto De Propris}
\affiliation{FINCA, University of Turku, Vesilinnantie 5, Turku, 20014, Finland}
\alsoaffiliation{Department of Physics and Astronomy, Botswana International University of Science and Technology, Private Bag 16, Palapye,
Botswana}

\author{Steven Bamford}
\affiliation{School of Physics and Astronomy, University of Nottingham, University Park, Nottingham NG7 2RD, UK}

\author{Karen Masters}
\affiliation{Departments of Physics and Astronomy, Haverford College, Lancaster Avenue, Ardmore, PA, 19041 USA}

\author{Maria Stone}
\affiliation{FINCA, University of Turku, Vesilinnantie 5, Turku, 20014, Finland}

\author{Tim Hardin}
\affiliation{University of Louisville, Department of Physics and Astronomy, 102 Natural Science Building, 40292 KY Louisville, USA.}

\author{Mike Walmsley}
\affiliation{Dunlap Institute for Astronomy and Astrophysics, University of Toronto, 50 St. George Street, Toronto, ON M5S 3H4, Canada}

\author{Jochen Liske}
\affiliation{Hamburger Sternwarte, Universit\"at Hamburg, Gojenbergsweg 112, 21029 Hamburg, Germany}

\author{S M Rafee Adnan}
\affiliation{University of Louisville, Department of Physics and Astronomy, 102 Natural Science Building, 40292 KY Louisville, USA.}

\keywords{
galaxies: structure < Galaxies	 Remove
galaxies: statistics < Galaxies	 Remove
galaxies: spiral < Galaxies	 Remove
galaxies: elliptical and lenticular, cD < Galaxies	 Remove
galaxies: bulges < Galaxies
}

\begin{document}

\begin{abstract}
Galaxy Zoo is an online project to classify morphological features in extra-galactic imaging surveys with public voting. In this paper, we compare the classifications made for two different surveys, the Dark Energy Spectroscopic Instrument (DESI) imaging survey and a part of the Kilo-Degree Survey (KiDS), in the equatorial fields of the Galaxy And Mass Assembly (GAMA) survey. Our aim is to cross-validate and compare the classifications based on different imaging quality and depth. 

We find that generally the voting agrees globally but with substantial scatter i.e. substantial differences for individual galaxies. There is a notable higher voting fraction in favor of ``smooth'' galaxies in the DESI+\rev{{\sc zoobot}} classifications, most likely due to the difference between imaging depth. DESI imaging is shallower and slightly lower resolution than KiDS and the Galaxy Zoo images do not reveal details such as disk features \rev{and thus are missed in the {\sc zoobot} training sample}. \rev{We check against expert visual classifications and find good agreement with KiDS-based Galaxy Zoo voting.}

We reproduce the results from Porter-Temple+ (2022), on the dependence of stellar mass, star-formation, and specific star-formation on the number of spiral arms. This shows that once corrected for redshift, the DESI Galaxy Zoo and KiDS Galaxy Zoo classifications agree well on population properties. The zoobot cross-validation increases confidence in its ability to compliment Galaxy Zoo classifications and its ability for transfer learning across surveys.
\end{abstract}

\section{Introduction}

Identifying galaxy morphological features remains something where the human eye is unsurpassed. Despite great improvements in machine learning techniques, the subtleties of various galaxy morphological properties remain best identified by experts or groups of people. Arguably, this makes these classifications prone to human error and biases. To remedy biases and scale up to the kind of imaging surveys now practical, the Galaxy Zoo project was conceived \citep{Lintott08, Willett13}. By involving the wider public in classifying galaxies with a series of specific questions, one can both gain morphological classifications and estimate their biases and uncertainties. 

Galaxy Zoo has had phenomenal success based on the work of thousands of volunteers working in tandem with professional astronomers. 
The results range from those that challenge our understanding of galaxy formation \citep[e.g.][]{Masters11, Smethurst15,Smethurst22, Walmsley19, Walmsley21, Walmsley21c} to a wealth of rare objects \citep[e.g.][]{Keel13,Keel22,Keel23}.

Morphological classifications are useful on their own but increase dramatically in utility when combined with other information, specifically Spectral Energy Distribution (SED) fits to multi-wavelength data and spectroscopic redshifts. The Galaxy and Mass Assembly \citep[GAMA,][]{Driver09,Driver21} survey is an excellent example of such a possible use. By combining {\sc MAGPHYS} SED information from GAMA with specific Galaxy Zoo classifications for the survey area, subtle differences in specific star-formation rates with a morphological feature can be identified \citep[e.g.][]{Porter-Temple22,Porter23,Smith21}. The improvements in accuracy, in both morphology from voting and inferred properties from the SED fit, for a statistical and complete samples allow for such more subtle relations to be revealed.

\begin{figure*}[htbp]
    \centering
    \includegraphics[width=0.8\textwidth]{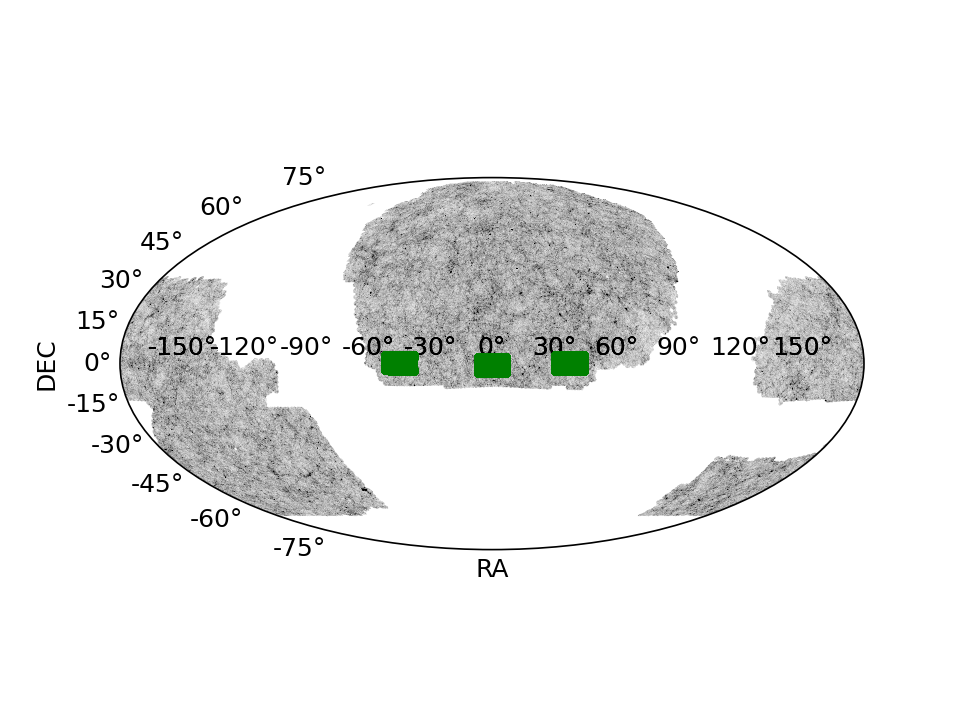}
    \caption{The relative coverage of the DESI-based Galaxy Zoo \protect\citep[black points][]{Walmsley23} and the Galaxy And Mass Assembly (GAMA) equatorial fields (green) in a Molleweide projection. There is full coverage in both DESI and KiDS for the equatorial GAMA fields. }
    \label{f:radecmap}
\end{figure*}

Our motivation to construct a GAMA Galaxy Zoo catalogue is therefore threefold: 
a) to further explore morphology relations with other inferred properties for this highly complete sample, 
b) to cross-examine morphology measures in two iterations of the Galaxy Zoo based on two different imaging surveys of the same galaxies (GAMA+KiDS and DESI Legacy Survey, Figure \ref{f:radecmap}) \rev{but completely different approaches (direct votes vs voting+{\sc zoobot} machine learning predictions)}, and 
c) provide a training catalogue for machine learning experiments for both researchers and students. 

\begin{figure*}[htbp]
    \centering
    \includegraphics[width=0.9\textwidth]{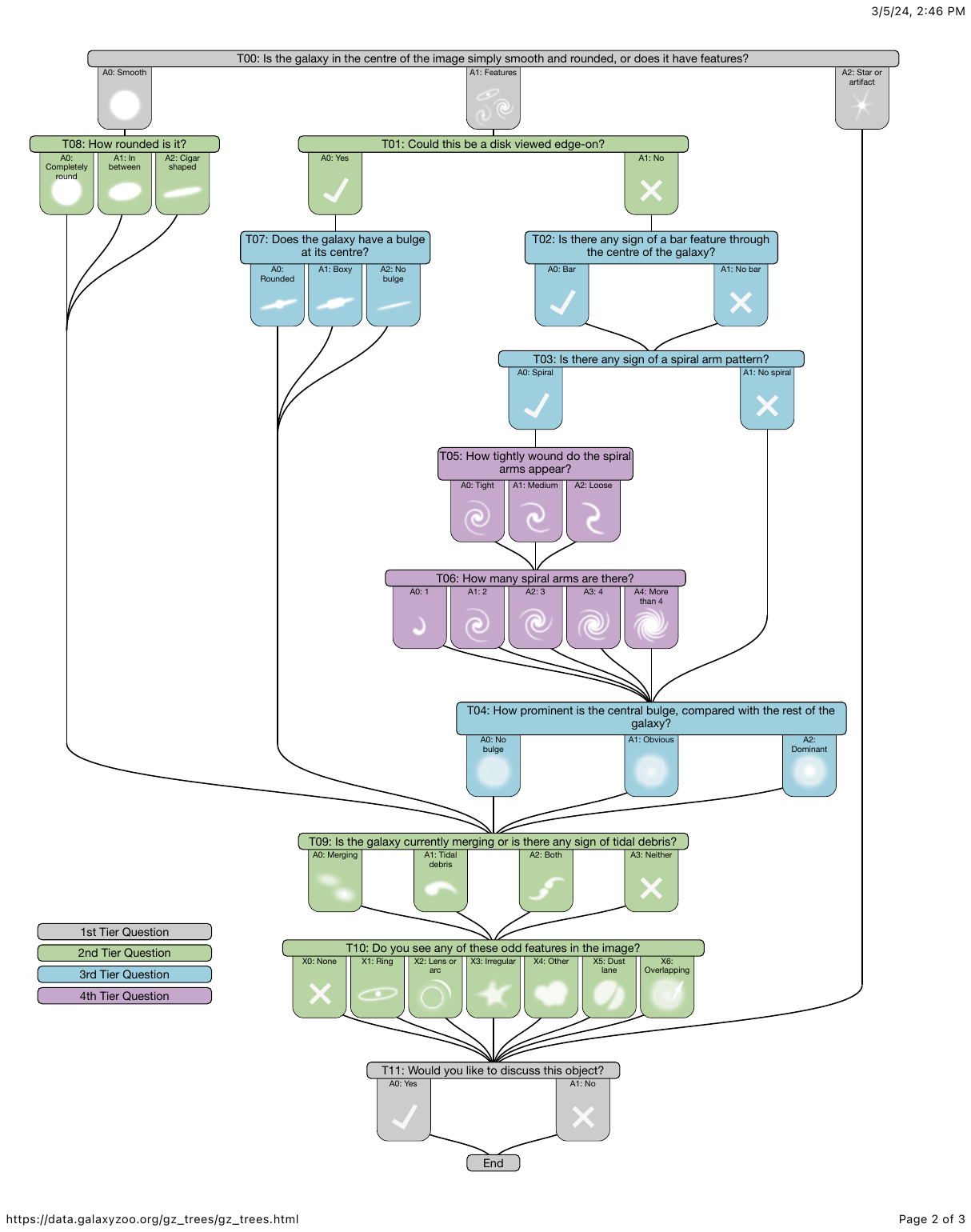}
    \caption{The flowchart of questions for volunteers for the KiDS-GZ database. Questions are listed in Table \ref{t:questions}}
    \label{f:flowchart}
\end{figure*}

The GAMA-KiDS Galaxy Zoo (KiDS-GZ) and the DESI-LS Galaxy Zoo (DESI-GZ) classifications are fortuitously based on mostly the same classification flow diagram for Galaxy Zoo Dr4 (Figure \ref{f:flowchart}), with the number of options for three questions changing slightly. This means the same questions were answered by the volunteers but based on different images. 
\rev{An additional difference between the two Galaxy Zoo catalogues is that the DESI-LS classifications catalogue is that it is based on {\sc zoobot} machine learning classifications trained on the Galayx Zoo classifications of all of DESI LS imaging. This study is meant to be a useful cross check between raw voting and a machine-learning extrapolated one.}

This paper is organized as follows: 
In Section \ref{s:data} we discuss the origin of the KiDS and DESI surveys Galaxy Zoo databases, as well as the parent GAMA survey catalogue. 
Section \ref{s:data-quality} briefly describes the differences in input imaging data quality from these surveys. 
In Section \ref{s:voting} we directly compare the reported voting fractions for each of the questions in Figure \ref{f:flowchart}, discussing patterns and their possible explanations. 
Section \ref{s:PTcomp} uses the DESI Galaxy Zoo voting matched with the GAMA {\sc MAGPHYS} catalogue to reproduce the results from \cite{Porter-Temple22} as an test of how well DESI-GZ can be used in GAMA. 
In Section \ref{s:discussion}, we discuss the benefits and drawbacks of using either Galaxy Zoo catalogue for future morphological studies, and 
Section \ref{s:conclusions} lists our conclusions from this comparison. 


\section{Data}
\label{s:data}

We use two iterations of the Galaxy Zoo catalogues: one based on KiDS images, made for the GAMA collaboration, and one made based on the DESI imaging survey as a larger Galaxy Zoo effort \rev{and then extrapolated using {\sc zoobot} \citep{Walmsley23}}. The relative coverage of these surveys, DESI and GAMA, are shown in Figure \ref{f:radecmap}. Ancillary data such as stellar mass estimates and redshifts are from the public GAMA-IV catalogues \citep[DR4][]{Driver22}. The other catalogues (target and {\sc MAGPHYS}) are part of GAMA DR2 release and available through the GAMA website.


\subsection{GAMA-KiDS Galaxy Zoo}

For GAMA, we use three sources: the GAMA survey itself, the KiDS survey for imaging, and finally the catalogue with voting from the Galaxy Zoo. 

\subsubsection{GAMA} \label{gama}
GAMA is a combined spectroscopic and multi-wavelength imaging survey designed to study spatial structure in the nearby ($z < 0.25$) Universe on kpc to Mpc scales \citep[see][for an overview]{Driver09, Driver22}. The survey, after completion of phase 2 \citep{Liske15}, consists of three equatorial regions each spanning 5 deg in Dec and 12 deg in RA, centered in RA at approximately 9h (G09), 12h (G12) and 14.5h (G15) and two Southern fields, at 02h (G02) and 23h (G23). The three equatorial regions, amounting to a total sky area of 180 deg$^2$, were selected for this study. For the purpose of visual classification, 49,851 galaxies were selected from the equatorial fields with redshifts $z<0.15$ (see below).   The GAMA survey is $>$95\% redshift complete to r $<$ 19.8 mag in all three equatorial regions \citep{Driver22}. We use the  {\sc magphys} SED fits data-products \citep{Driver18} from the third GAMA data-release \citep[DR3,][]{Baldry18}. The GAMA Galaxy Zoo voting catalogue is slated to be part of the rolling DR4 \citep{Driver22}.

\subsubsection{KiDS} \label{kids}

The Kilo Degree Survey \citep[KiDS,][]{de-Jong13,de-Jong15,de-Jong17,Kuijken19} is an optical wide-field imaging survey with the OmegaCAM camera at the VLT Survey Telescope. It has imaged 1350 deg$^2$ in four filters (u g r i). The core science driver is mapping the large-scale matter distribution in the Universe, using weak lensing shear and photometric redshift measurements. Further science cases include galaxy evolution, Milky Way structure, detection of high-redshift clusters, and finding rare sources such as strong lenses and quasars.
KiDS image quality is typically 0\farcs7 resolution (for sdss-r) and depths of 23.5, 25, 25.2, and 24.2 magnitude for i, r, g and u, respectively. This imaging was the input for the Galaxy Zoo citizen science classifications \citep[see also][]{Kelvin18}. 

\subsubsection{KiDS Galaxy Zoo} 
\label{s:kids-gz}

Information on galaxy morphology is based on the GAMA-KiDS Galaxy Zoo classification \citep{Lintott08, Kelvin18}. 
GAMA-KiDS Galaxy Zoo data was initially provided by Lee Kelvin and Steven Bamford (\textit{private communication}, 2019). Further details of this project may be found in \cite{Holwerda19}, \cite{Porter-Temple22}, and \cite{Porter23}.
Briefly, Red-Green-Blue (RGB) cutouts were constructed from KiDS g-band and r-band imaging with the green channel as the mean of these. KiDS cutouts were introduced to the classification pool and mixed in with the ongoing classification efforts. For the Galaxy Zoo classification, 49,851 galaxies were selected from the equatorial fields with redshifts $z < 0.15$. Galaxy Zoo provided a monumental effort with almost 2 million classifications received from over 20,000 unique users over the course of 12 months. 
This classification has been used by the GAMA team to identify dust lanes in edge-on galaxies \citep{Holwerda19},  searches for strong lensing galaxy pairs \citep{Knabel20}, the morphology of green valley galaxies \citep{Smith22} and void galaxies \citep{Porter23} and to link star-formation properties to the number of spiral arms \citep{Porter-Temple22}.
In this paper we use the visual classifications of disk  galaxies from the Galaxy Zoo project; the full decision tree for the GAMA-KiDS Galaxy Zoo project is shown in Figure \ref{f:flowchart}. When talking about this catalogue, we use KiDS-GZ. 

\subsection{DESI Galaxy Zoo}
\label{s:desi-gz}

\cite{Walmsley23} presents the galaxy zoo classifications based on three colour imaging of the Dark Energy Spectroscopic Instrument (DESI) survey \citep{Dey19}. DESI requires images to target its spectroscopic fibers; these are primarily provided by the DESI Legacy Surveys (DESI-LS). These images are shallower than KiDS and often lower resolution (poorer seeing) which varied from field to field \citep{Dey19}. The images from DESI-LS were converted into \rev{Red-Green-Blue (RGB)} images for processing in the Galaxy Zoo classification pipeline.8 

\rev{\cite{Walmsley23} provide user-friendly catalogues for use in future endeavours such as these. There are two catalogues: one with the Galaxy Zoo classifications over all of DESI-LS. These were used to train {\sc zoobot} \citep{Walmsley23c}, a deep-learning tool, to predict voting fractions for all of DESI-LS sources. This is the catalogue shown in Figure \ref{f:radecmap}. Because DESI-LS covers a wide area, the number of galaxies with actual voting in the GAMA footprint is small. We use the {\sc zoobot} predictions catalogue for this reason to directly compare to the KiDS-GZ voting. }

\rev{The {\sc zoobot} \citep{Walmsley23c} is a deep-learning tool that can be trained (and retrained) to predict Galaxy Zoo volunteer voting fractions based on RGB and grayscale images. The code is publicly available here: \url{https://github.com/mwalmsley/zoobot}. 
{\sc zoobot} is based on the Keras Python function \textit{EfficientNetB0}, an image classifier using an optimized Convolutional Neural Network structure \citep{Tan19}. 
The catalogues are publicly available here: \url{https://zenodo.org/records/8360385}. } 
\rev{Because this is a prediction, not actual voting records for these galaxies, the voting fraction for example never reaches 100\% in the DESI-GZ catalogue we are using (see section 4.1 on question T00). }



\subsection{Catalogue Matching}

We match the catalogues from KiDS-GZ and the DESI-GZ using their positions (the {\sc match\_coords\_sky} algorithm in {\sc astropy}) within a match radius of two arcseconds (the width of the GAMA spectroscopic fibre). We opt to match only the GAMA equatorial fields (G09, G12, G15) and not G02 which also has Galaxy Zoo voting information because the equatorial fields have the most ancillary catalogs in the GAMA Data-releases (e.g. MAGPHYS SED fits). Previous work using the KiDS-GZ \citep{Holwerda19,Knabel20,Knabel23,Porter-Temple22,Porter23} all use only the equatorial fields. 

Matching the GAMA targeting catalogue and DESI-GZ results in some 35k sources in common. The G02 field has 10k sources in common but these are not used in the rest of the paper. The DESI-GZ catalogue matches to GAMA are released as part of this paper with GAMA CATAID added for future GAMA archival use. 

\begin{table}[htbp]
    \centering
    \begin{tabular}{l l l}
         &  \\
            & KiDS      & DESI Legacy Survey \\ 
\hline
sdss-g      & 25.2      & 23.48-23.72 \\ 
sdss-r      & 25        & 23.27-22.87 \\ 
sdss-i      & 23.5      & 22.22-22.29 \\ 
         \hline
    \end{tabular}
    \caption{Survey Depths of KiDS and the DESI Legacy Survey. The depth of DESI-LS depends strongly on which sub-survey is used but it is at least a magnitude shallower than KiDS.  }
    \label{t:depths}
\end{table}

\section{Data Quality}
\label{s:data-quality}
\subsection{Image Quality}

The KiDS-GZ classifications are based on RGB images with the green filter constructed from red and blue, as the green filter (sdss-r) had not been observed for all targets yet at the beginning of the Galaxy Zoo iteration. The DESI-LS survey was obtained on a variety of telescopes but with the same filter set as SDSS and KiDS. Survey depths are summarized in Table \ref{t:depths}. Spatial resolution varies similarly to depth (0\farcs6--1\farcs2) for DESI-LS while KiDS is more consistent with an average PSF of 0\farcs7.
On the whole, the DESI-GZ classifications are based on more complete information, e.g. including a green filter, and more area, \rev{but these are extrapolated using {\sc zoobot}}. The GAMA-KiDS imaging is higher resolution and deeper in the two filters used. 

The physical resolution in the respective surveys depends on the seeing and distance. In previous work \citep{Smith22b,Porter-Temple22,Porter23,Holwerda22}, samples were limited in redshift to keep sensitivity to smaller morphological features the same across the studied sample. In the case of the KiDS-GZ for example, a redshift limit of z=0.08 corresponds to a physical resolution of $\sim$1 kpc, a key resolution for morphology \citep{Lotz04}.


\begin{figure}[htbp]
    \centering
    \includegraphics[width=\textwidth]{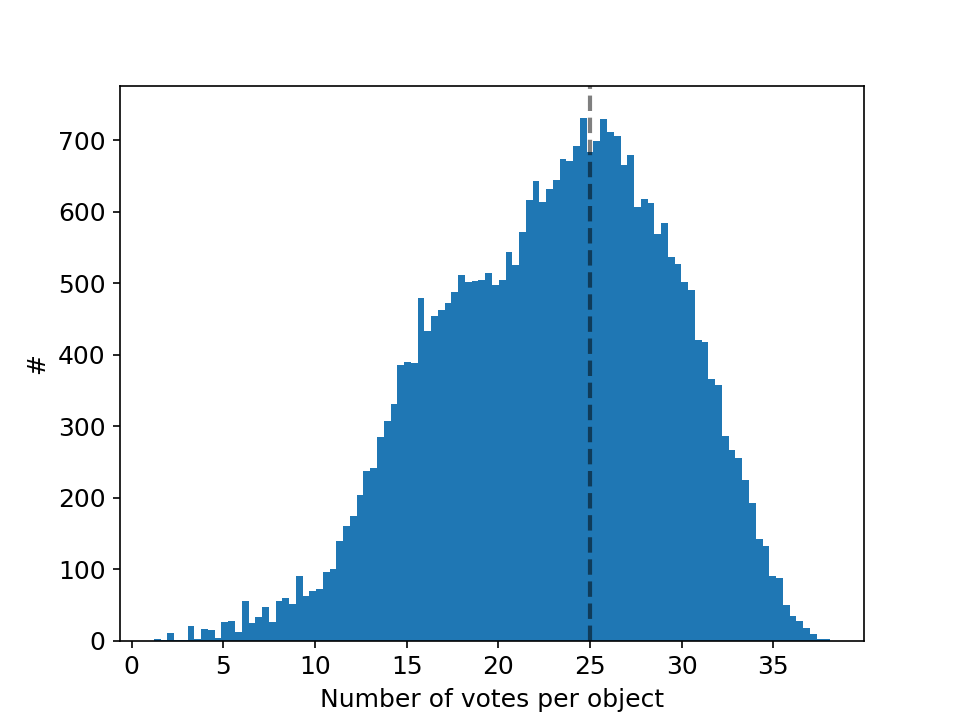}
    \caption{The number of votes in question T00 (see Figure \ref{f:flowchart}), which is asked for every galaxy. The mode for KiDS-GZ is 25 classifications per object (dashed line) and the mean 23.}
    \label{f:hist_classifiers}
\end{figure}

\subsection{Number of Classifiers}

The histogram of the number of classifiers for each KiDS-GZ is shown in Figure \ref{f:hist_classifiers}. The mean number is 36 classifiers for the DESI-GZ \rev{training sample}. The number of classifiers is lower for KiDS-GZ, leading to possibly larger scatter in the voting fractions. We note that the DESI-GZ catalogue, the KiDS-GZ is compared against is the extrapolated one by {\sc zoobot}.  


The trade off between the two Galaxy Zoo iterations is that the KiDS-GZ imaging may be deeper and likely higher resolution, but the DESI-GZ has a higher number of votes going into \rev{individual sources of the large training sample which is then generalized to all the DESI-GZ catalogue \citep{Walmsley23}.}
If the {\em zoobot} classifier is trained well on the DESI-GZ, the accuracy should be equal to or surpassing the lower number of classifiers on the deeper KiDS data.

\begin{table*}[htbp]
    \centering
    \begin{tabular}{l|l r r}
    \hline
Number & Question & \# Options & \\
& & KiDS & DESI \\
\hline
T00: & Is the galaxy in the centre of the image simply smooth and rounded, or does it have features? & 3 & \\
T01: & Could this be a disk viewed edge-on? & 2 & \\
T02: & Is there any sign of a bar feature through the centre of the galaxy? & 2 & 3\\
T03: & Is there any sign of a spiral arm pattern? & 2 & \\
T04: & How prominent is the central bulge, compared with the rest of the galaxy? & 3 & 5\\
T05: & How tightly wound do the spiral arms appear? & 3 & \\
T06: & How many spiral arms are there? & 5 & 6 \\
T07: & Does the galaxy have a bulge at its centre? & 3 & \\
T08: & How rounded is it (the galaxy)? & 3 & \\
T09: & Is the galaxy currently merging or is there any sign of tidal debris? & 4 & \\
T10: & Do you see any of these odd features in the image? & 7 & \\ 
    \hline
    \end{tabular}
    \caption{The questions in the Galaxy Zoo 4th iteration (KiDS-GZ and DESI-GZ). The number of options are given. The number of options for T02, T04 and T06 were higher for DESI-GZ.}
    \label{t:questions}
\end{table*}

\begin{figure*}[htbp]
    \centering
    \includegraphics[width=\linewidth]{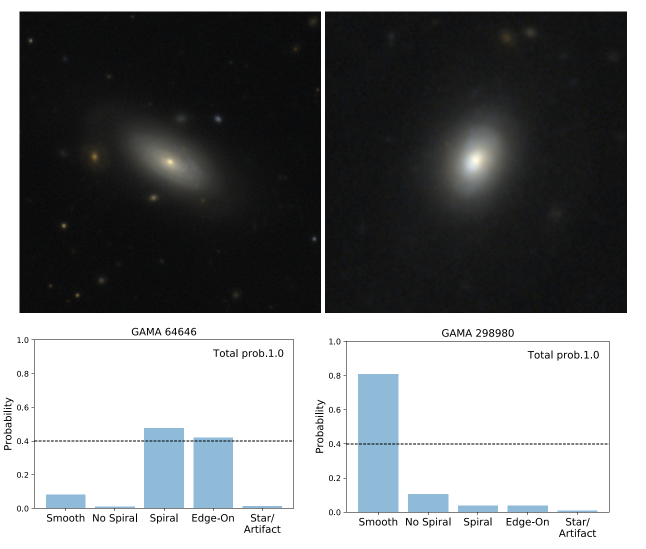}
    \caption{Two examples of KiDS galaxies with their respective voting fractions for questions T00, T01 and T03. From the thesis of David Henry William Glass \citep{glassthesis}.}
    \label{f:KiDS:voting}
\end{figure*}

\section{Voting Comparison}
\label{s:voting}
Figure \ref{f:flowchart} shows the flowchart of questions for the GAMA-KiDS classifications. We will compare the fraction of votes for each question. 

Figure \ref{f:KiDS:voting} shows two KiDS images of ETGs examined by KiDS-GZ classifiers, and histograms of probabilities (multiples of vote fractions within relevant questions) for five endpoints within the flowchart in Figure \ref{f:flowchart} \citep{glassthesis}. 
The chosen endpoints, Smooth (T00:A0), 
Edge-On (T00:A1 $ \times $ T01:A0), 
Spiral (T00:A1 $ \times $ T01:A1 $ \times $ T03:A0), 
No\_Spiral (T00:A1 $ \times $ 
T01:A1 $ \times $ T03:A1) 
and Star/Artifact (T00:A2), form a complete set horizontally across 
Question T00 and probabilities therefore sum to 1. The threshold for morphology selection was the dominant probability above 0.4, to ensure a leading selection with only one other choice close behind, as is the case for GAMA64646.

T01-T03 are binary choices, allowing for easier comparisons using either fraction. In the case of multiple options, the voting fraction for each needs to be compared. We note that DESI-GZ and KiDS-GZ had a different number of options for T02 and T04, where the DESI-GZ questionnaire had more options. In these cases, we compare the same worded answer, with much of the difference due to a difference in choice. 
We compare voting fractions since the number of votes is different for KiDS-GZ and DESI-GZ as the retirement criterion --the point in voting where the image was not shown to new classifiers-- was set differently between these. Most usage focuses on voting fractions to identify features, not absolute or calibrated numbers of votes.

\subsection{T00: Is the galaxy in the centre of the image simply smooth and rounded, or does it have features? }

This is the first question encountered to separate those galaxies that are mostly featureless (elliptical/spheroidal) and artifacts from those with a lot of substructures (Figure \ref{f:flowchart}). This is a key question as the voter is not shown the remaining detailed questions if they do not mark the objects as ``having features''.

\rev{The voting in DESI-GZ does not reach 100\% as does KiDS-GZ. This is an artifact of the {\sc zoobot} predictions. In the Galaxy Zoo voting in the DESI-LS, 100\% fractions do happen. }

\begin{figure}[htbp]
    \centering
    \includegraphics[width=\textwidth]{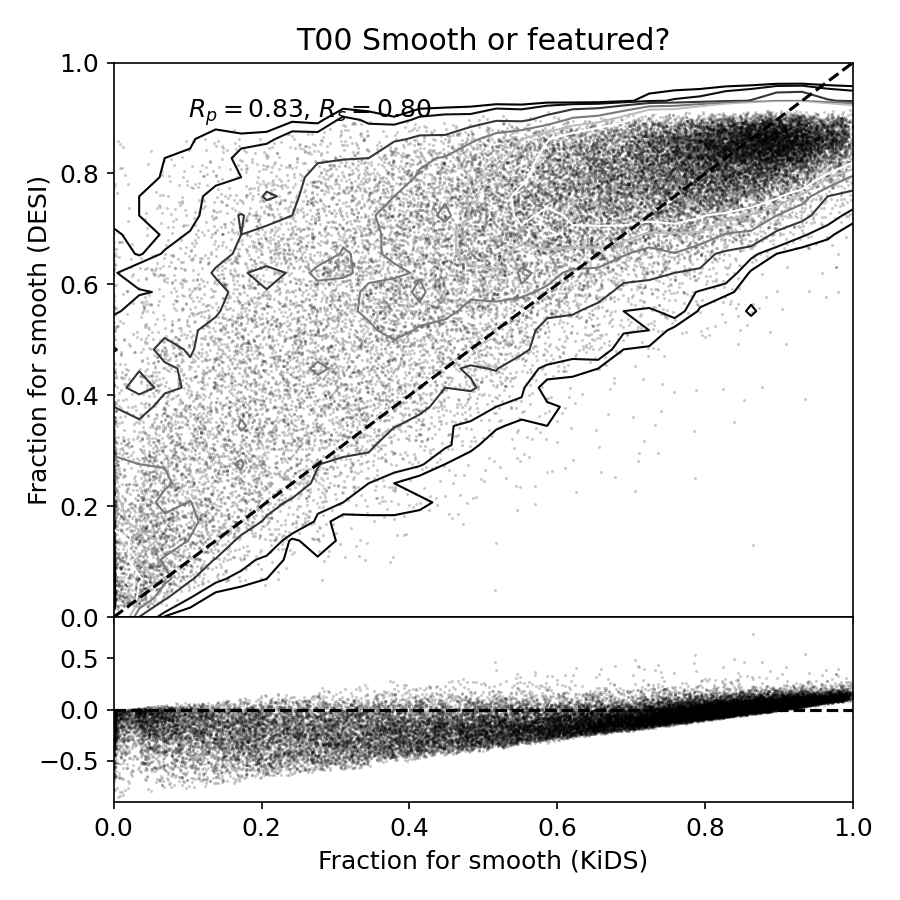}
    \caption{The fractions of votes in question T00 (Table \ref{t:questions}) in favor of these galaxies to be ``smooth''. On the x-axis is the voting fraction for the KiDS-GZ and on the y-axis the DESI-GZ voting fraction \rev{predictions by the {\sc zoobot}} (top panel) or the difference between the two (bottom panel). Each point is a galaxy in common between the DESI-GZ and the KiDS-GZ. T00 is one of three questions that is asked for each object. Contours are drawn at 5,10,25,50,75, and 100 densities. 
    $R_P$ and $R_S$ are the Pearson and Spearman ranking in the relation, respectively. }
    \label{f:T00}
\end{figure}



Figure \ref{f:T00} shows the comparison between the KiDS-GZ and the DESI-GZ \rev{{\sc zoobot} predicted} voting fractions for the galaxies in common. This is one of the questions where we have answers for all the galaxies. T00 is the first question to be answered and thus always presented to all classifiers. A notable behavior is that the fraction of positive votes for smooth in the DESI-GZ voting is typically higher than the KiDS-GZ one. This means Galaxy Zoo volunteers voted for galaxies to be smooth more in DESI-GZ than in GAMA-KiDS. \rev{This DEIS-GZ voting is then reflected in the {\sc zoobot} training sample.}
This is likely due to the deeper and/or higher resolution in the GAMA-KiDS images as volunteers are able to identify more galaxies with structure of some kind. 

\rev{It also highlights a difference the {\sc zoobot} prediction makes on a voting fraction: there are no 100\% voting fractions in the DESI-GZ. There are some in the KiDS-GZ but experience has shown these often to be the result of a single vote, especially for questions that depend on a previous choice (T01--T08). We leave the 100\% fractions in our KiDS-GZ catalogue for now but often rejected these in analyses for this reason. An alternate approach is to renormalize these question with all the volunteers considering this object, not the number that answered the question.}

\begin{figure}[htbp]
    \centering
    \includegraphics[width=\textwidth]{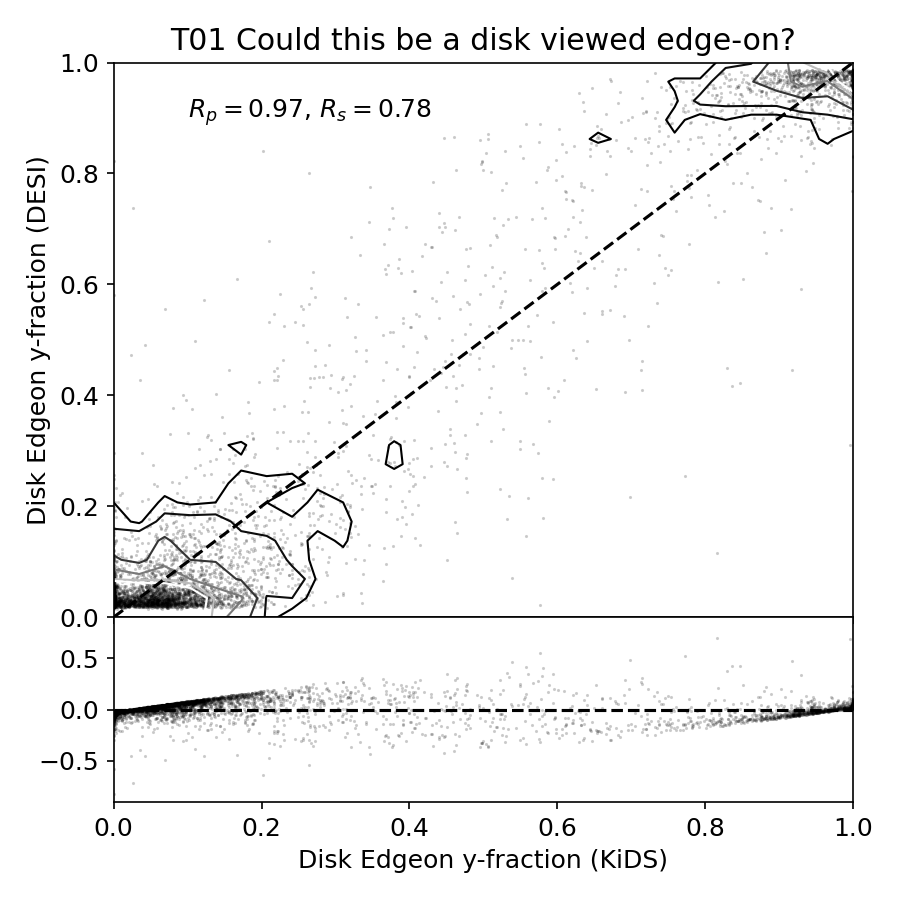}
    \caption{The fractions of votes in question T01 (Table \ref{t:questions}) in favor of these galaxies to be ``edge-on''. On the x-axis is the voting fraction for the KiDS-GZ and on the y-axis the DESI-GZ \rev{{\sc zoobot} predicted} voting fraction (top panel) or the difference between the two (bottom panel). Values of exact 0 and exactly 1 are ignored as these likely indicate single (erroneous) vote counts. }
    \label{f:T01}
\end{figure}



\subsection{T01: Could this be a disk viewed edge-on? }

Rather than an axis ratio, this is a direct question if this is possibly an edge-on disk galaxy. 
As a binary question (only two answers possible; yes or no), we only need to examine one voting fraction because the other answer's listed fraction is the inverse of the first answer. 
This question immediately follows voting in favor of this galaxy having ``features''. The comparison sample is therefore smaller because a voting fraction is only recorded if someone in both groups of volunteers has voted on this galaxy to have features. There are two populations visible in Figure \ref{f:T01}, galaxies with a high fraction in both DESI-GZ \rev{{\sc zoobot} predictions} and KiDS-GZ in favor of the edge-on perspective and a larger group with a small fraction ($f<0.2$) in favor of the edge-on perspective. 

The voter fractions in Figure \ref{f:T01} agree quite well in case of the ``edge-on'' question. This might make a sample of edge-on galaxies selected by this plot quite robust. 

\subsection{T02: Is there any sign of a bar feature through the centre of the galaxy? }

\begin{figure}[htbp]
    \centering
    \includegraphics[width=\textwidth]{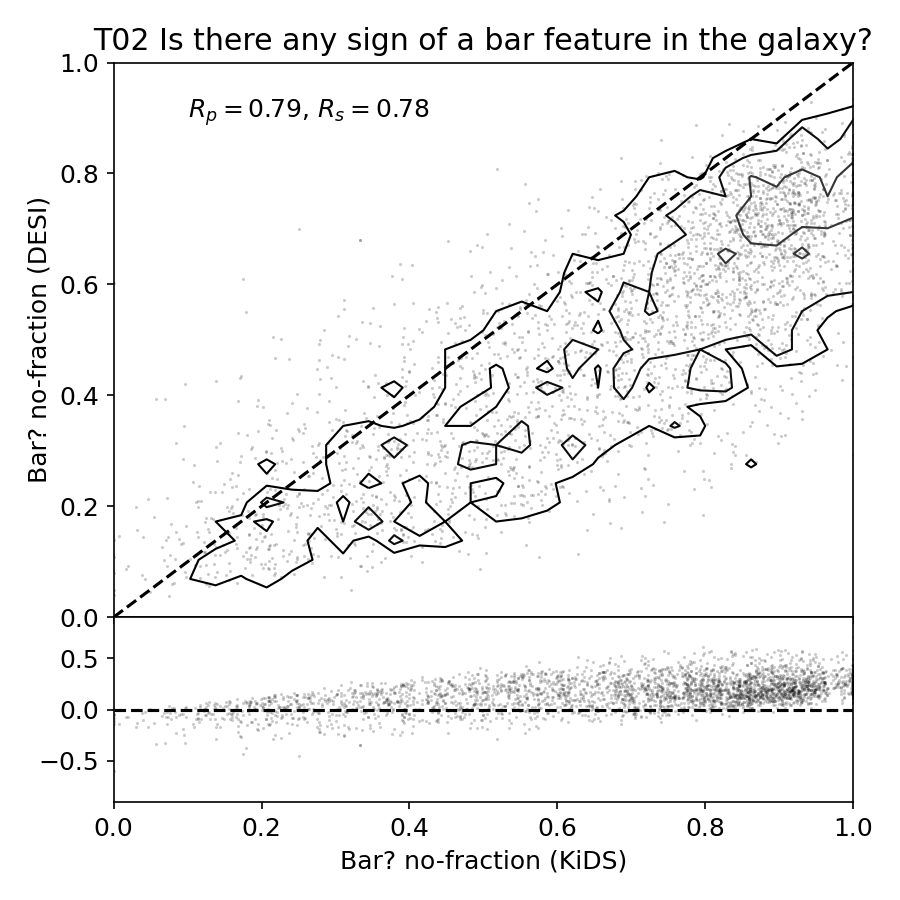}
    \caption{The fractions of votes in question T02 (Table \ref{t:questions}) in favor of these galaxies to \textit{not} have a bar. On the x-axis is the voting fraction for the KiDS-GZ and on the y-axis the DESI-GZ \rev{{\sc zoobot} predicted} voting fraction (top panel) or the difference between the two (bottom panel). This is one of the questions that changed between KiDS-GZ and DESI-GZ with the addition of ``strong'' and ``weak'' in the latter. }
    \label{f:T02}
\end{figure}

This is a y/n question in the KiDS-GZ iteration, but one that was expanded to three options in DESI-GZ; none/weak/strong options for the bar. The fractions for ``no-bar'' are compared in Figure \ref{f:T02}. This is the only option common to both surveys. There is reasonable agreement between the two GZ iterations but the change in options could mean that the fraction for no-bar in DESI is lower as the option for ``weak bar'' is now available.

\begin{figure}[htbp]
    \centering
    \includegraphics[width=\textwidth]{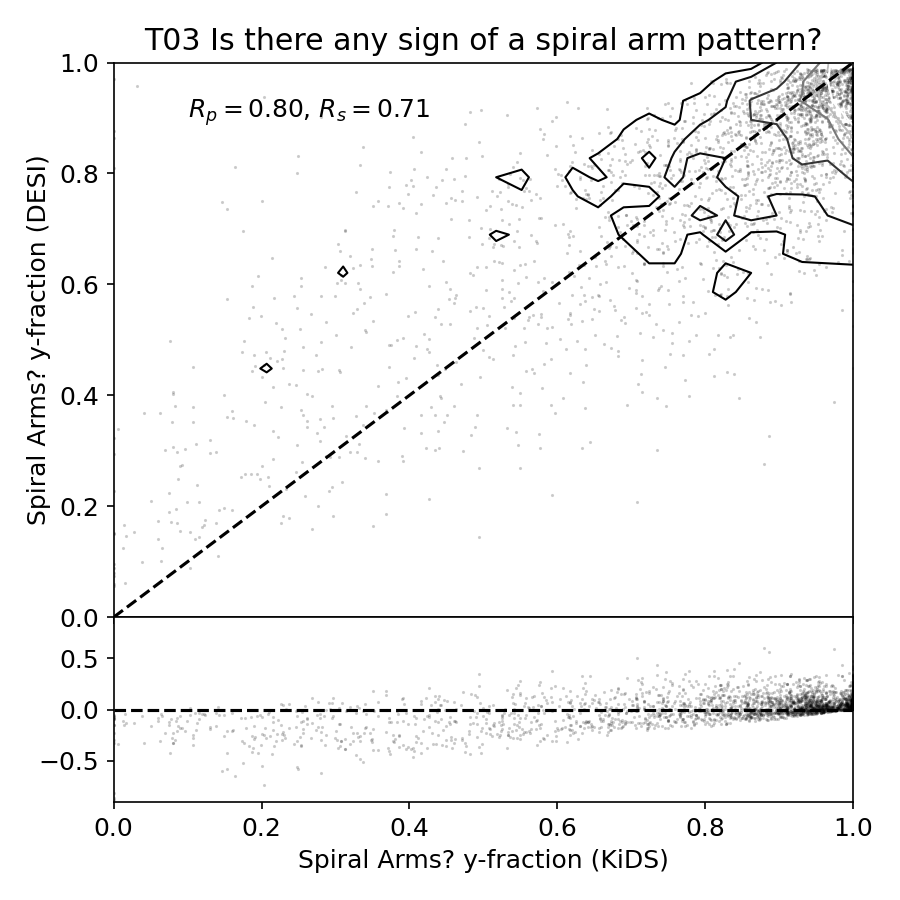}
    \caption{The fractions of votes in question T03 (Table \ref{t:questions}) in favor of these galaxies to have an identifiable spiral pattern. On the x-axis is the voting fraction for the KiDS-GZ and on the y-axis the DESI-GZ voting fraction (top panel) or the difference between the two (bottom panel).}
    \label{f:T03}
\end{figure}

\subsection{T03: Is there any sign of a spiral arm pattern? }

The third and last binary (y/n) question. The fractions of voting are shown in Figure \ref{f:T03}. This is a question only answered after T00 and T01, and therefore the comparison sample is once again smaller. There is a good agreement for high fractions of voting in both DESI-GZ \rev{{\sc zoobot} predictions} and KiDS-GZ. 



Glass et al. \citep[2024, {\em in prep.}][]{glassthesis} showed that KiDS-GZ vote fractions can be used successfully to identify and remove weak spiral galaxies in samples of smooth early-type galaxies derived from GAMA classifications \citep{Kelvin14, Moffett16a}. The GAMA classifications are based on SDSS imaging (catalogue VisualMorphologyv03), at lower resolution and depth than KiDS. For disc-like early-type galaxies (ellipticals) in their sample, 19\% were found to have weak spiral features using KiDS-GZ.


\begin{figure}[htbp]
    \centering
    \includegraphics[width=\textwidth]{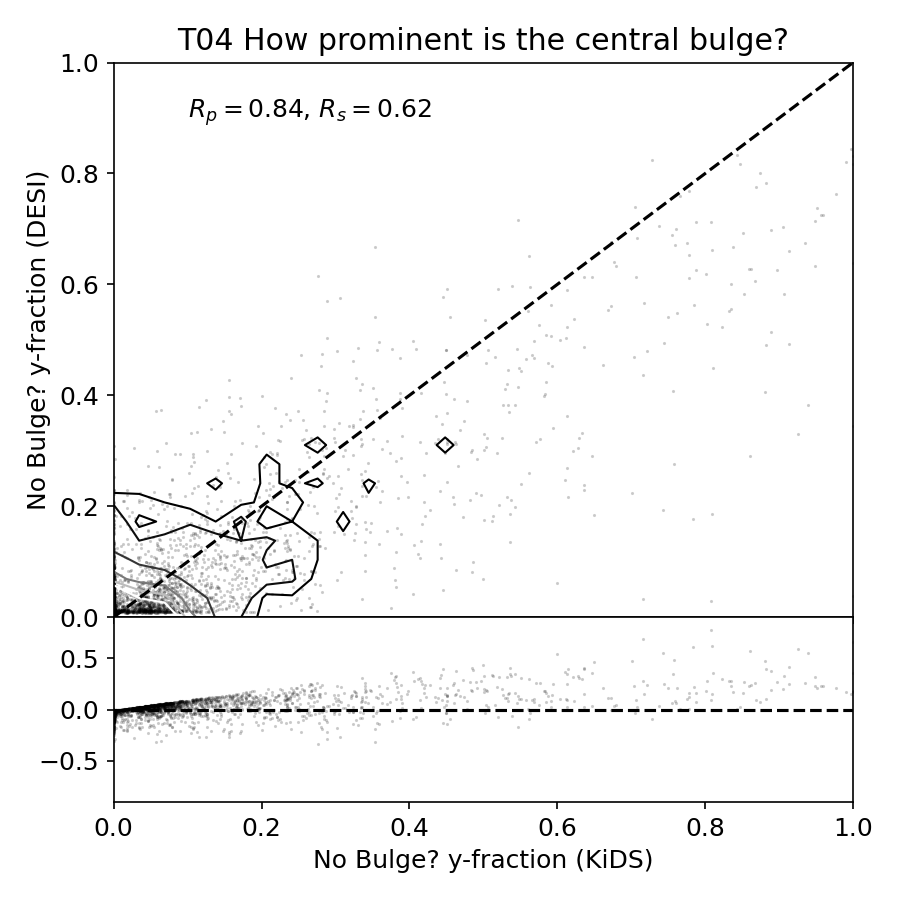}
    \caption{The fractions of votes in question T04 (Table \ref{t:questions}) in favor of these galaxies to have a ``prominent'' central bulge. On the x-axis is the voting fraction for the KiDS-GZ and on the y-axis the DESI-GZ \rev{{\sc zoobot} predicted} voting fraction (top panel) or the difference between the two (bottom panel).}
    \label{f:T04}
\end{figure}

\subsection{T04: How prominent is the central bulge, compared with the rest of the galaxy? }

This is a question with three options in the KiDS-GZ but five options in the DESI-GZ \rev{and the subsequent {\sc zoobot} predictions}. In principle, it could be reduced to a binary one (evidence of a bulge y/n?) similar to T07. 

\rev{Figure \ref{f:T04} shows the ``no bulge'' voting fractions for KiDS-GZ and DESI-GZ. This is the answer both Galaxy Zoo questions have in common. The agreement with other answers for this question is quite poor. The most likely explanation is that the DESI-GZ offered more options to classify. Many of the prominent bulges would have gotten votes for ``large'' bulge instead. Other possible explanations include the difference in depth between DESI-GZ and KiDS-GZ and that the RGB images were constructed differently in each survey. The fact that ``no bulge'' is consistent is encouraging. 
}


\begin{figure}[htbp]
    \centering
    \includegraphics[width=\textwidth]{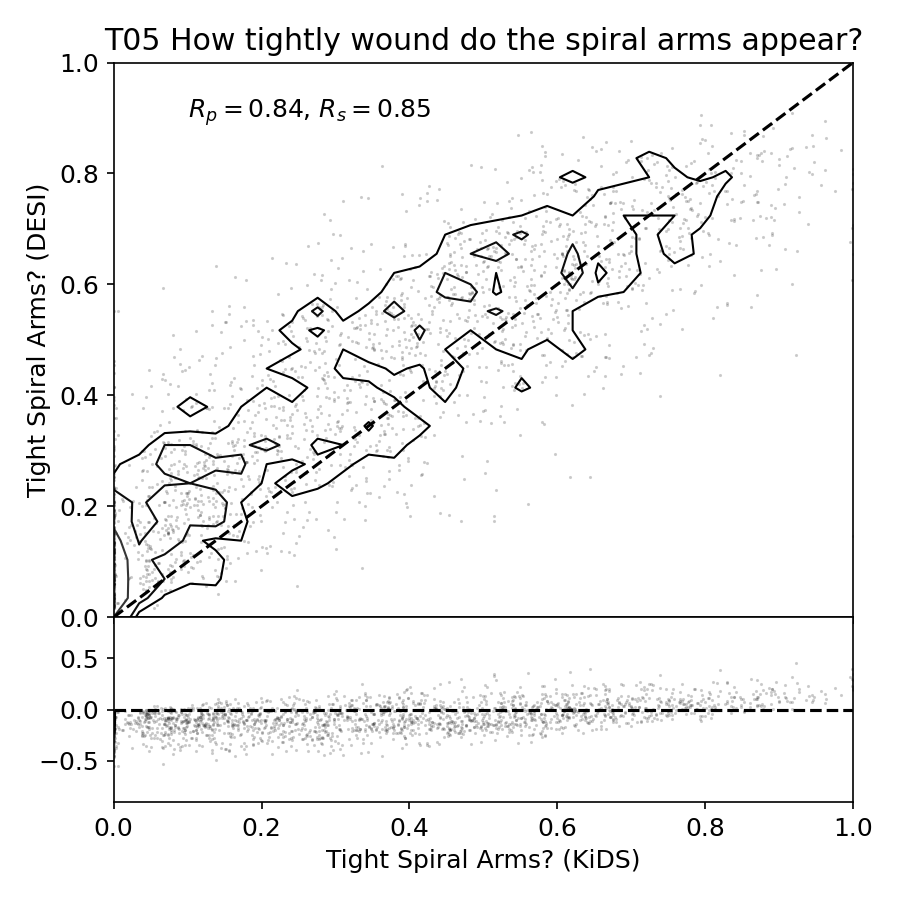}
    \caption{The fractions of votes in question T04 (Table \ref{t:questions}) in favor of these galaxies to have a ``tightly wound'' spiral arms. On the x-axis is the voting fraction for the KiDS-GZ and on the y-axis the DESI-GZ \rev{{\sc zoobot} predicted} voting fraction (top panel) or the difference between the two (bottom panel).}
    \label{f:T05}
\end{figure}

\subsection{T05: How tightly wound do the spiral arms appear? }

This question is only presented if the Galaxy Zoo volunteer answers affirmative to T03. The three answers are not easily reduced to a binary question. Figure \ref{f:T05} shows the voting fractions for the ``tightly wound'' answer compared between KiDS-GZ and DESI-GZ \rev{{\sc zoobot} predictions} . 

The reasonably good agreement between the answer fractions for KiDS-GZ and DESI-GZ \rev{{\sc zoobot} predictions} is reassuring that once spiral arm features are identified in either survey, there is good agreement on their appearance.

\begin{figure*}[htbp]
    \centering
    \includegraphics[width=0.24\textwidth]{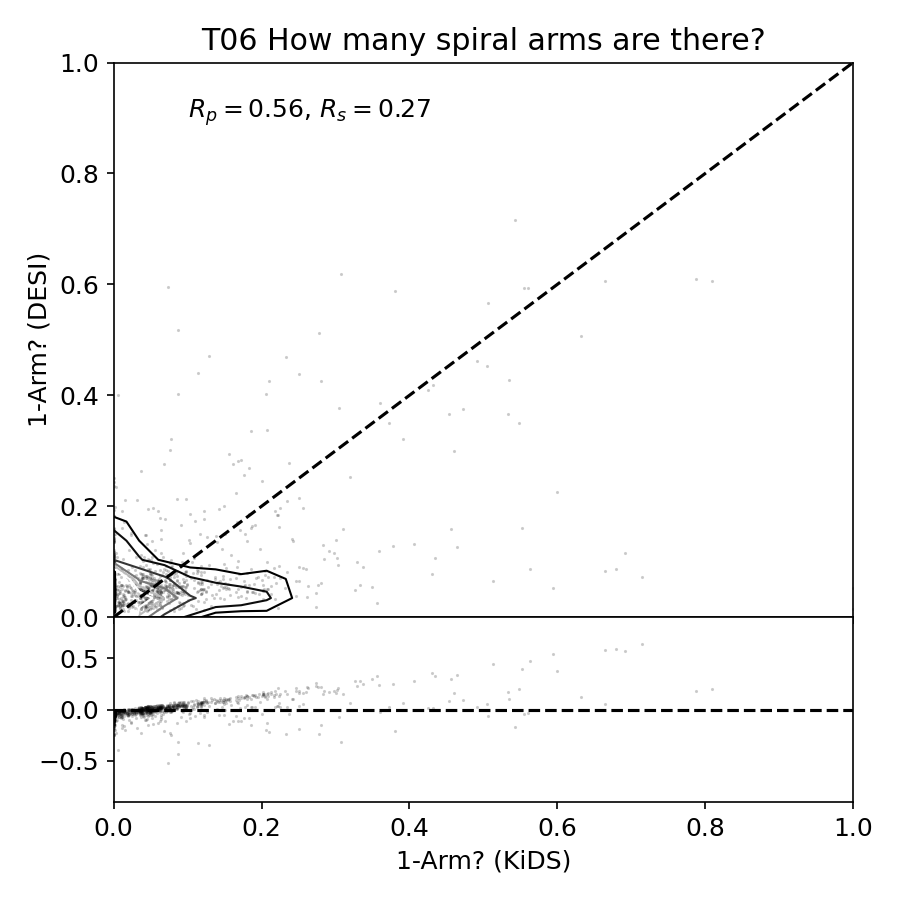}
    \includegraphics[width=0.24\textwidth]{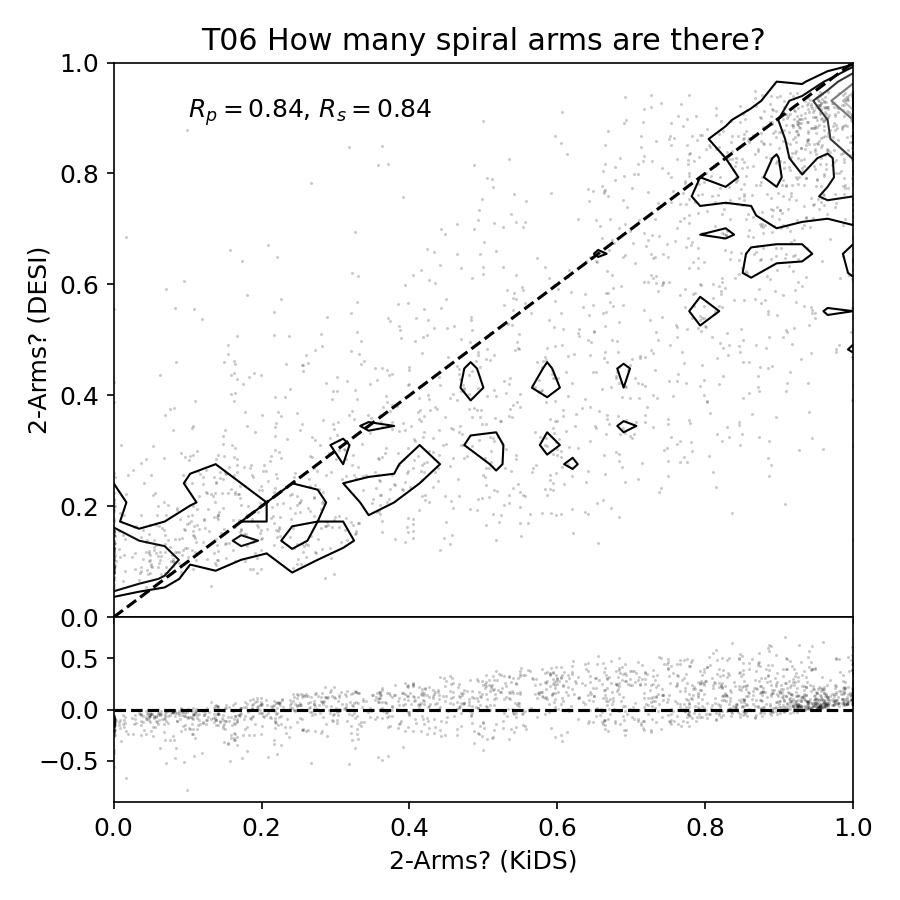}   \includegraphics[width=0.24\textwidth]{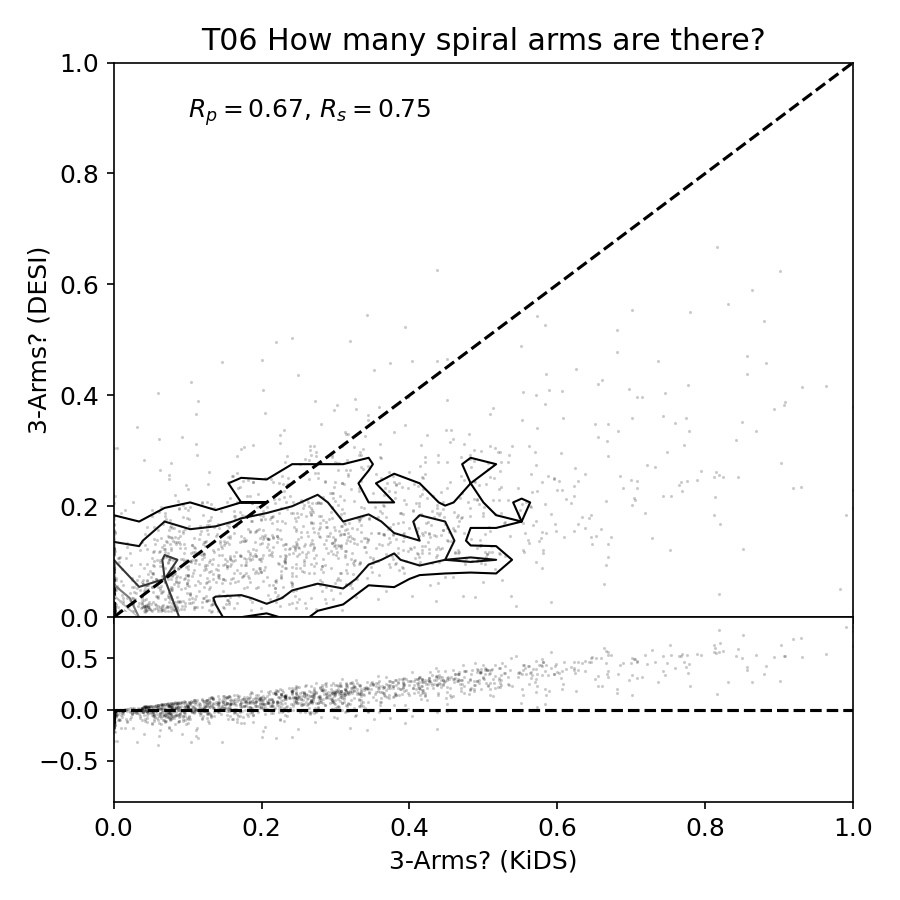}    \includegraphics[width=0.24\textwidth]{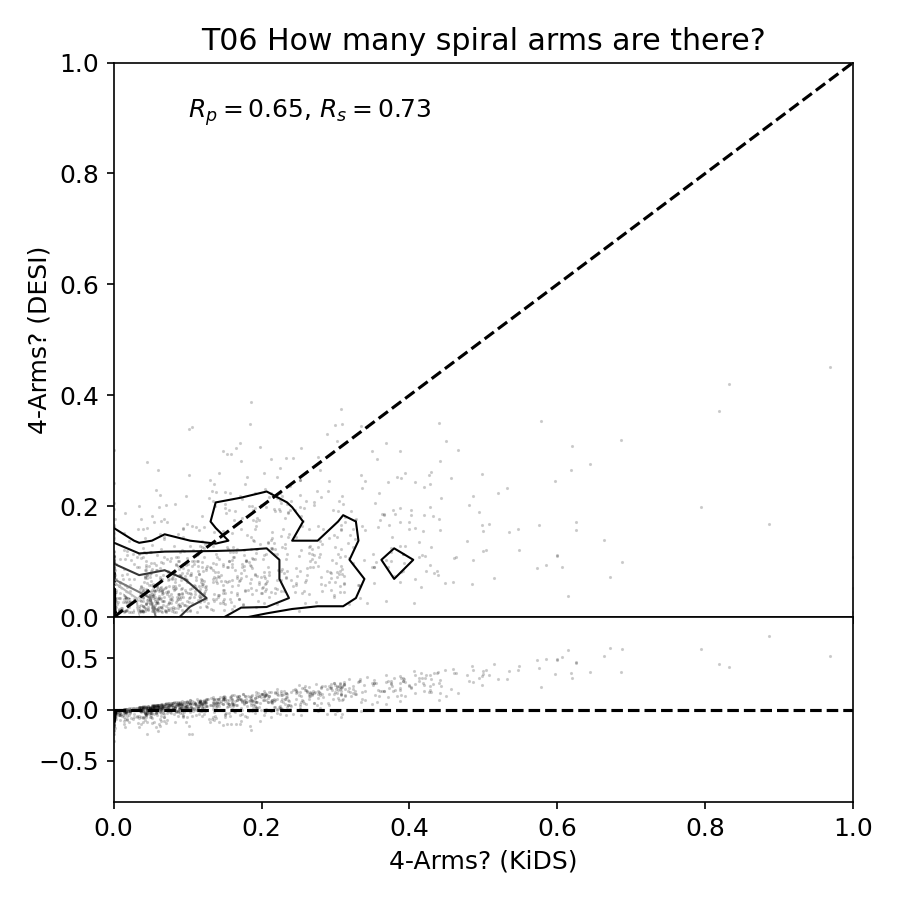}    
    \caption{The fractions of votes in question T06 (Table \ref{t:questions}) on the number of spiral arms, one to four from left to right. On the x-axis is the voting fraction for the KiDS-GZ and on the y-axis the DESI-GZ \rev{{\sc zoobot} predicted} voting fraction (top panel) or the difference between the two (bottom panel).}
    \label{f:T06}
\end{figure*}

\subsection{T06: How many spiral arms are there? }

This is the question on which \cite{Porter-Temple22} focused for their study of star-formation and stellar mass properties. The option ``more than 4'' for the number of spiral arms is functionally a vote for a flocculent spiral. In principle, this question could be reduced to a binary one of ``grand design spiral'' or ``flocculent'' but this has not been used. \cite{Porter-Temple22} found the vast majority of objects in their sample ($\rm log(M_*/M_\odot > 9$, $z<0.08$) for completeness and resolution reasons) to mostly consist of 2-armed spirals with much smaller but statistically significant numbers of the other categories. 

Figure \ref{f:T06} shows the voting fractions for KiDS-GZ and DESI-GZ \rev{{\sc zoobot} predictions} for all four distinct answers in question T06. The best agreement is for 2-armed spirals (the most common kind). One-armed spirals are agreed upon with a low fraction, and a similar but more diffuse version of that pattern repeats for 3-armed spirals. There is a noticeable trend with both 3- and 4-armed spirals where there is a higher voting fraction in favor of these in the KiDS-GZ compared to the DESI-GZ \rev{{\sc zoobot} predictions}.
There is an additional answer possible in DESI-GZ (``cannot tell'') which was not an option in KiDS-GZ. 
This all suggests the threshold for identifying a larger number of arms may have to be set to a lower fraction in the DESI-GZ \rev{{\sc zoobot} predictions}. There are different ways to ensure a galaxy has a certain number of spiral arms. One could require a simple majority ($>$20\% for a given option within the five options in T06) or an overwhelming consensus ($>$50\%), depending on how certain one wants to be of the selected sample. The former can still be a close call (all options have almost 1/5th of the vote), while the latter is unambiguous but lower statistics.
This question is relevant for any comparison to \cite{Porter-Temple22} or \cite{Hart17}: ``how well do volunteers agree on the number of spiral arms in a galaxy'' (see \ref{s:PTcomp}). 




\begin{figure}[htbp]
    \centering
    \includegraphics[width=\textwidth]{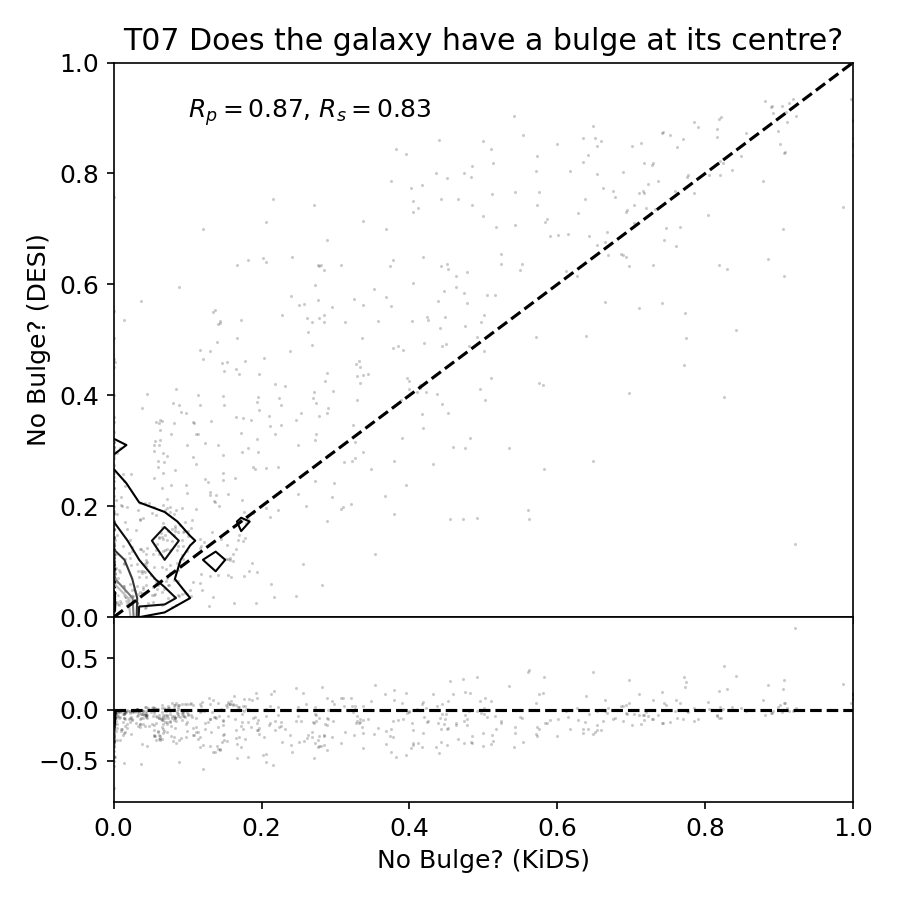}
    \caption{The fractions of votes in question T04 (Table \ref{t:questions}) in favor of these edge-on galaxies to have no bulge. On the x-axis is the voting fraction for the KiDS-GZ and on the y-axis the DESI-GZ \rev{{\sc zoobot} predicted} voting fraction (top panel) or the difference between the two (bottom panel).}
    \label{f:T07}
\end{figure}

\subsection{T07: Does the galaxy have a bulge at its centre? }

This question is only answered if T01 is positive (view is edge-on). Three answers are possible (Figure \ref{f:flowchart}). In principle, this question can be reduced to a binary one (is there a bulge y/n?) by combining the voting of the first two options (``boxy'' and ``round''). Figure \ref{f:T07} shows the fraction of votes in favor of ``no bulge''. Generally speaking, the two Galaxy Zoo iterations, KiDS-GZ and DESI-GZ \rev{{\sc zoobot} predictions} are in broad agreement but with large scatter. 

\begin{figure*}[htbp]
    \centering
    \includegraphics[width=0.32\textwidth]{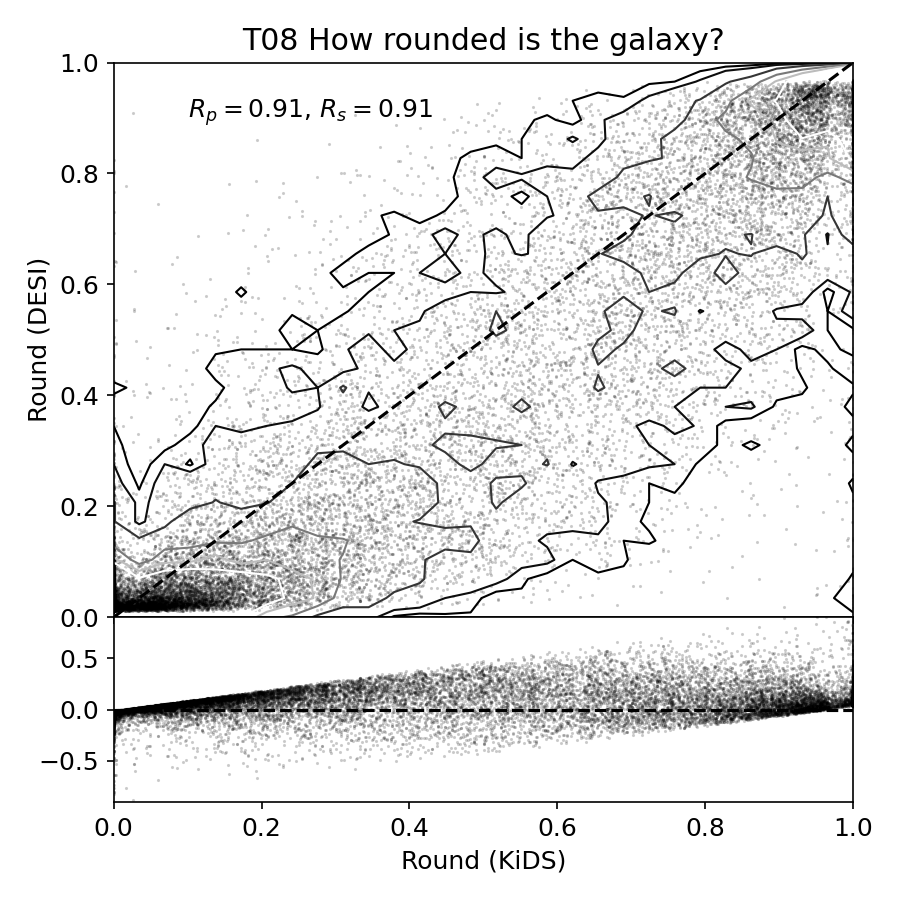}
    \includegraphics[width=0.32\textwidth]{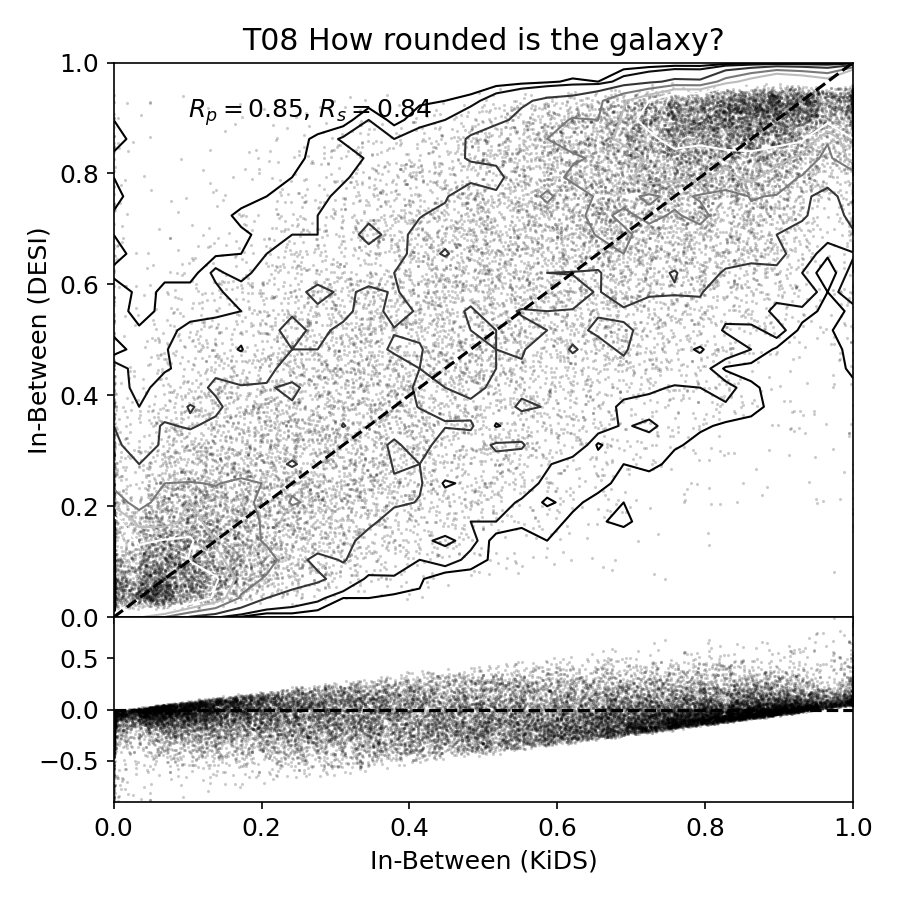}
    \includegraphics[width=0.32\textwidth]{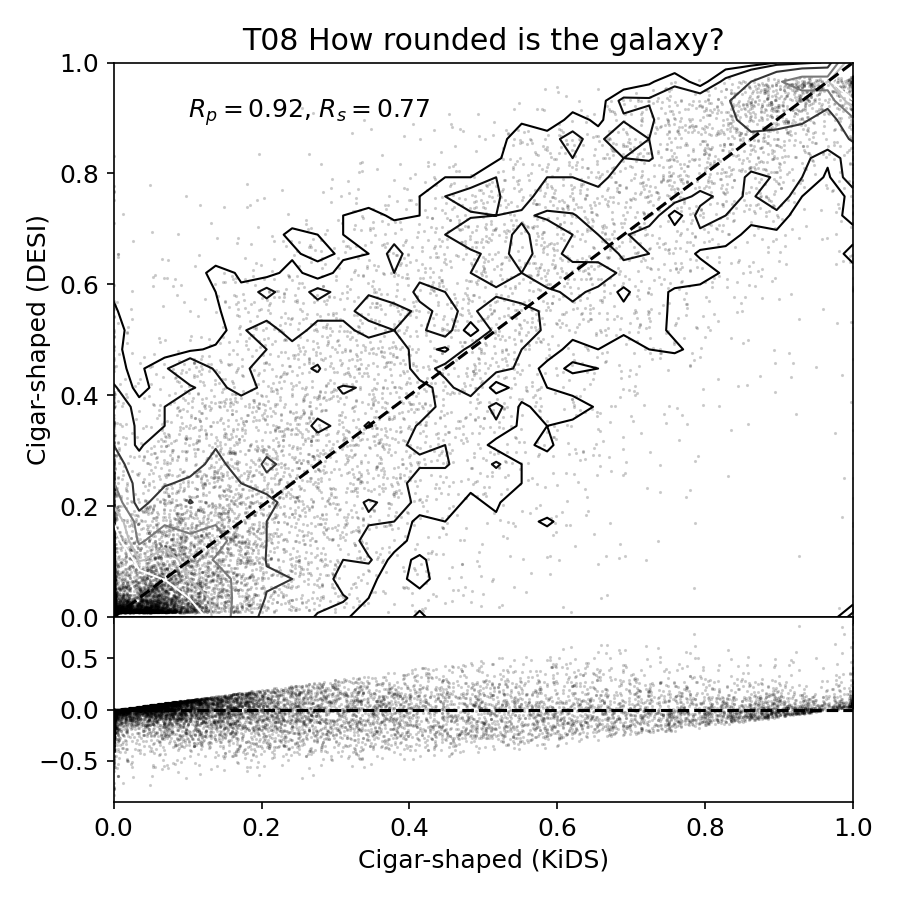}
    \caption{The fractions of votes in question T08 (Table \ref{t:questions}) in favor of smooth galaxies that appear round (left panel), in-between (middle panel), or cigar-shaped (right panel). On the x-axis is the voting fraction for the KiDS-GZ and on the y-axis the DESI-GZ \rev{{\sc zoobot} predicted} voting fraction (top panel) or the difference between the two (bottom panel).}
    \label{f:T08}
\end{figure*}

\subsection{T08: How rounded is it? }

This is the only dedicated question if the volunteer answers ``Smooth'' in T00. This is exclusively for elliptical/spheroidal galaxies. Figure \ref{f:T08} shows the voting fractions for all three options (round, in-between, and cigar) compared between the KiDS-GZ and DESI-GZ \rev{{\sc zoobot} predictions}. There is general agreement but with substantial scatter, to be expected for a slightly relative or subjective question.

\begin{figure*}[htbp]
    \centering
    \includegraphics[width=0.49\textwidth]{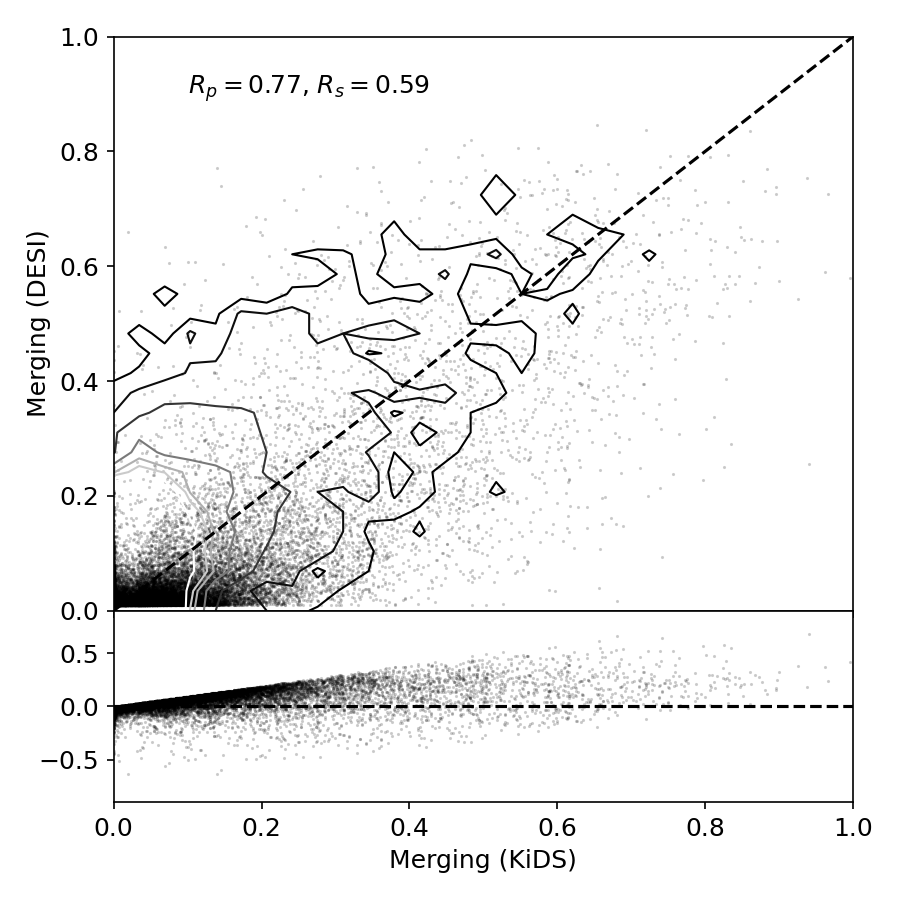}
    \includegraphics[width=0.49\textwidth]{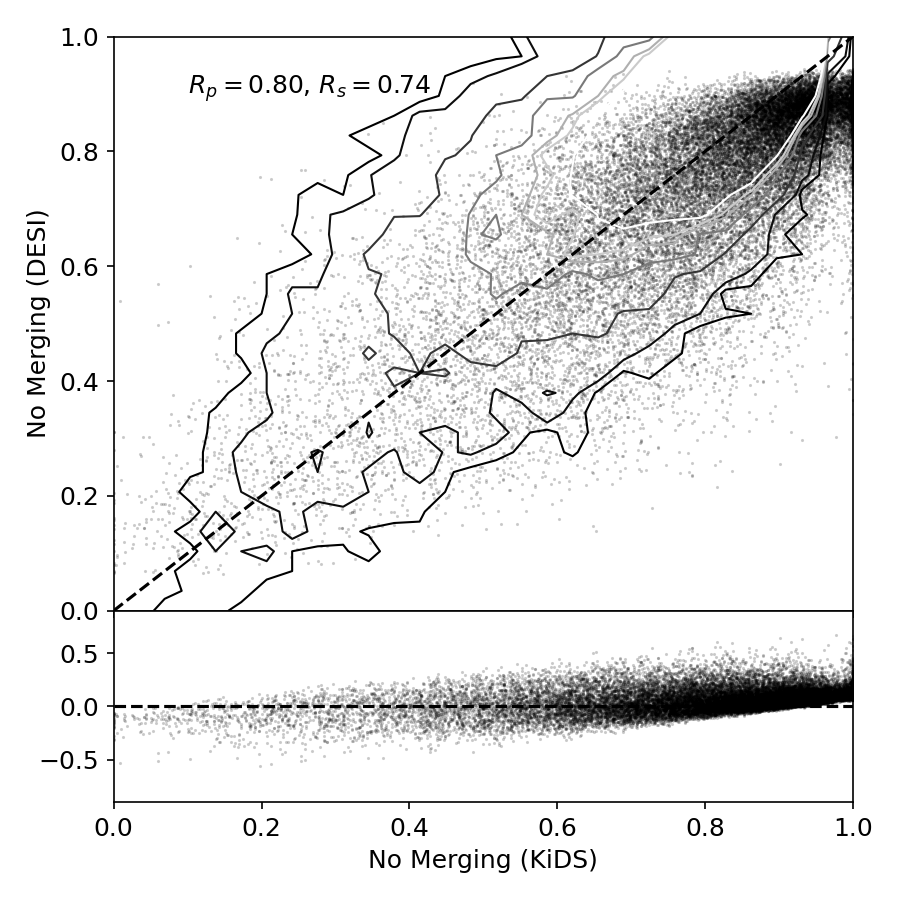}
    \caption{The fractions of votes in question T09 (Table \ref{t:questions}) in favor of an interaction or not. On the x-axis is the voting fraction for the KiDS-GZ and on the y-axis the DESI-GZ \rev{{\sc zoobot} predicted} voting fraction (top panel) or the difference between the two (bottom panel).}
    \label{f:T09}
\end{figure*}

\subsection{T09: Is the galaxy currently merging or is there any sign of tidal debris? }

This is a question with four possible answers but could be reduced to a single ``signs of interaction y/n?'' by combining the first three answers to compare against the ``none'' voting fraction. This was the usage in \cite{Porter23} for void galaxies. 
The middle options between ``merging'' and ``none'' in this question are the only change between KiDS-GZ and DESI-GZ \rev{{\sc zoobot} predictions}. The middle options are ``tidal debris'' and ``both'' (meaning the galaxies show both signs of merging and tidal debris) for the KiDS-GZ and the middle options in DESI-GZ \rev{{\sc zoobot} predictions} are ``minor'' and ``major'' indicating the relative ratio of the galaxies in minor/major interaction. 

Figure \ref{f:T09} shows the comparison between the two answers the GZ iterations have in completely common (i.e. the answer is phrased the same). This is a question that is asked for all galaxies, regardless of T01 so the comparison has more statistics. Generally, there is a reasonable agreement (high fraction of no-merger in both iterations) but especially at lower fractions (more ambivalence), the scatter is higher. There is a better agreement on no-merger than on merging since the other options could draw votes away depending on the image depth (i.e. a tidal feature is visible in KiDS but not in DESI, \rev{either in the volunteer voting or the {\sc zoobot} predictions}).

\subsection{T10: Do you see any of these odd features in the image? } 

The final question is unique in that the answers are not mutually exclusive and one could vote for more than one of these features. For example, one could see an overlapping pair of galaxies and a prominent dust lane visible. Whether or not it is clear to each volunteer that multiple answers are allowed is not clear. 

This question was not included in the data-release by \cite{Walmsley23} and we do not include the comparison here. The question was undoubtedly asked but it would be difficult to inter-compare with likely low statistics as these are relatively infrequently occurring phenomena. \rev{{\sc zoobot} predictions for these are difficult for the same reasons.}

\begin{table}[htbp]
    \centering
    \begin{tabular}{l|l r r}
    \hline
    \hline
    Question & Pearson R (p-value) & Spearman R (p-value) & Figure\\
    \hline 
T00 & 0.83 (0.00) & 0.80 (0.00) & Fig. \ref{f:T00}\\ 
T01 & 0.97 (0.00) & 0.78 (0.00) & Fig. \ref{f:T01}\\ 
T02 & 0.79 (0.00) & 0.78 (0.00) & Fig. \ref{f:T02}\\ 
T03 & 0.80 (0.00) & 0.71 (0.00) & Fig. \ref{f:T03}\\ 
T04 & 0.84 (0.00) & 0.62 (0.00) & Fig. \ref{f:T04}\\ 
T05 & 0.84 (0.00) & 0.85 (0.00) & Fig. \ref{f:T05}\\ 
T06-1 & 0.56 (1.48E-169) & 0.27 (2.18E-35) & Fig. \ref{f:T06}\\ 
T06-2 & 0.84 (0.00) & 0.84 (0.00) & Fig. \ref{f:T06}\\ 
T06-3 & 0.67 (1.28E-275) & 0.75 (0.00) & Fig. \ref{f:T06}\\ 
T06-4 & 0.65 (3.28E-248) & 0.73 (0.00) & Fig. \ref{f:T06}\\ 
T07 & 0.87 (0.00) & 0.83 (1.58E-272) & Fig. \ref{f:T07}\\ 
T08-1 & 0.91 (0.00) & 0.91 (0.00) & Fig. \ref{f:T08}\\ 
T08-2 & 0.85 (0.00) & 0.91 (0.00) & Fig. \ref{f:T08}\\ 
T08-3 & 0.92 (0.00) & 0.91 (0.00) & Fig. \ref{f:T08}\\ 
T09-1 & 0.77 (0.00) & 0.59 (0.00) & Fig. \ref{f:T09}\\ 
T09-2 & 0.80 (0.00) & 0.74 (0.00) & Fig. \ref{f:T09}\\ 
    \hline
    \end{tabular}
    \caption{The questions in the Galaxy Zoo 4th iteration (GAMA-KiDS and DESI-LS \rev{{\sc zoobot} predictions}). The number of options are given. }
    \label{t:questions:stats}
\end{table}

\subsection{Correlation Metrics}

Figures \ref{f:T00} through \ref{f:T09} include the Pearson ($R_p$) and Spearman ($R_s$) rankings. These are summarized in Table \ref{t:questions:stats} including the p-values returned with each test. The Pearson ranking is an indication of how linear the relation between the two voting fractions is. The Spearman one is a ranking for a monotonous, but importantly not necessarily linear, relation between the two voting fractions. 

Most of the voting between KiDS-GZ and DESI-GZ \rev{{\sc zoobot} predictions} is highly correlated with rankings well above 0.8. The highest agreement is on whether this disk can be viewed edge-on (T01). This is reflected in Figure \ref{f:T01} with clusters at 0 and 1. 

\rev{The lower correlations are often for questions where either there were more options in one of the Galaxy Zoo iterations (e.g. T04) or a suspected dependence on surface brightness (e.g. T03), or both.}


Of the number of spiral arms, a subtle difference in surface brightness may make a difference. The agreement is the strongest for two arms, where the statistics are the highest. The agreement on mergers (T09) are surprisingly good because these are dominated by the answer that there is no evidence for an ongoing or past merger (Figure \ref{f:T09}).


\section{Other GAMA Visual Classifications}
\label{s:viscom}

Previous visual classifications of the GAMA galaxies include those by the GAMA team \citep{Driver22} and low-redshift quasar hosts \citep{Stone23}. These are visual classifications of the galaxy as a whole (Table \ref{t:vizclasses}). We compare the KiDS-GZ voting against these expert visual classifications. 

\begin{table*}[htbp]
    \centering
    \begin{tabular}{l l l}
NumClass & Class & Description \\
\hline
 & & \cite{Driver22} classifications \\
\hline
0 &  E     & Elliptical system with a single visual component   \\
1 &  cBD   & Two-component system with a compact high-surface brightness bulge \\
2 &  dBD   & Two-component system with a diffuse or extended bulge (or bulge complex) \\
3 &   D	   & Disk system with a single visual component \\
\hline
 & & \cite{Stone23} classifications \\
\hline
0 & -  & Early-type (E or S0) \\
1 & -  & Middle-type (Sa or Sb) \\
2 & -  & Late-type (Sc or later) \\
3 & - & Unknown types (possible merger)\\
\hline
    \end{tabular}
    \caption{Visual classification schemes used in \cite{Driver22} and \cite{Stone23} for GAMA galaxies.}
    \label{t:vizclasses}
\end{table*}


\begin{figure}[htbp]
    \centering
    \includegraphics[width=\linewidth]{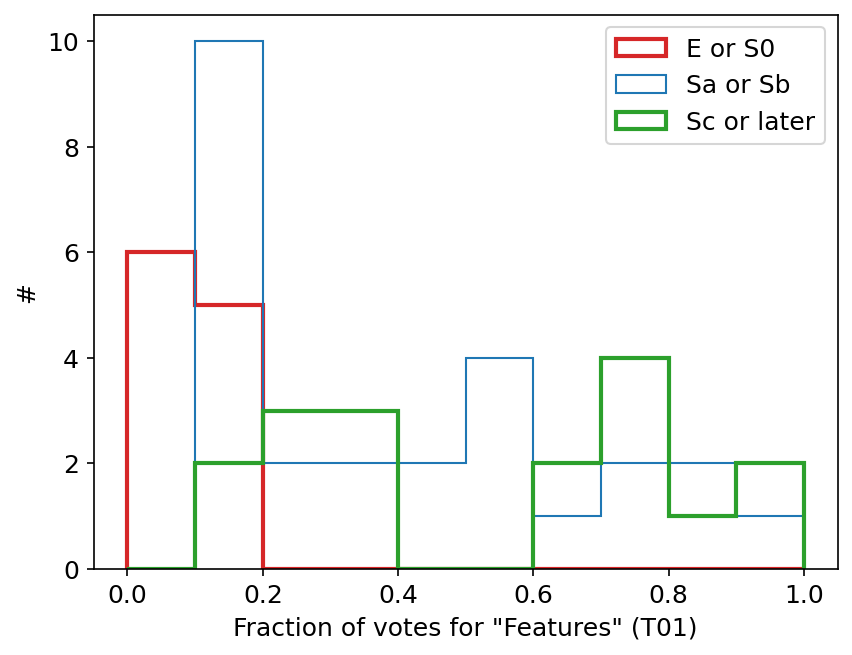}
    \caption{The histogram of classified galaxies as a function of the voting fraction of T00 question ``smooth or featured?" in the DESI-GZ. The \cite{Stone23} classifications show a low fraction for ellipticals (E-S0) and higher voting fraction  for disk galaxies (either ``Sa or Sb'' and ``Sc or later''). This makes the Galaxy Zoo classifications consistent with the expert visual assessment from \cite{Stone23}.}
    \label{f:viz:stone}
\end{figure}

Figure \ref{f:viz:stone} shows the overlap (58 galaxies out of 205) with the DESI-GZ sample and the one from \cite{Stone23} for quasar host galaxies and question T00, the most numerous and relevant one for the morphology of the galaxy as a whole. Early type galaxies (E or S0) show a low fraction of ``features" votes, later types, i.e. disk dominated classes have higher fractions of votes in favor of ``features''. This makes the Galaxy Zoo classifications consistent with the expert visual assessment in \cite{Stone23} of quasar host galaxies.

\begin{figure}[htbp]
    \centering
    \includegraphics[width=\linewidth]{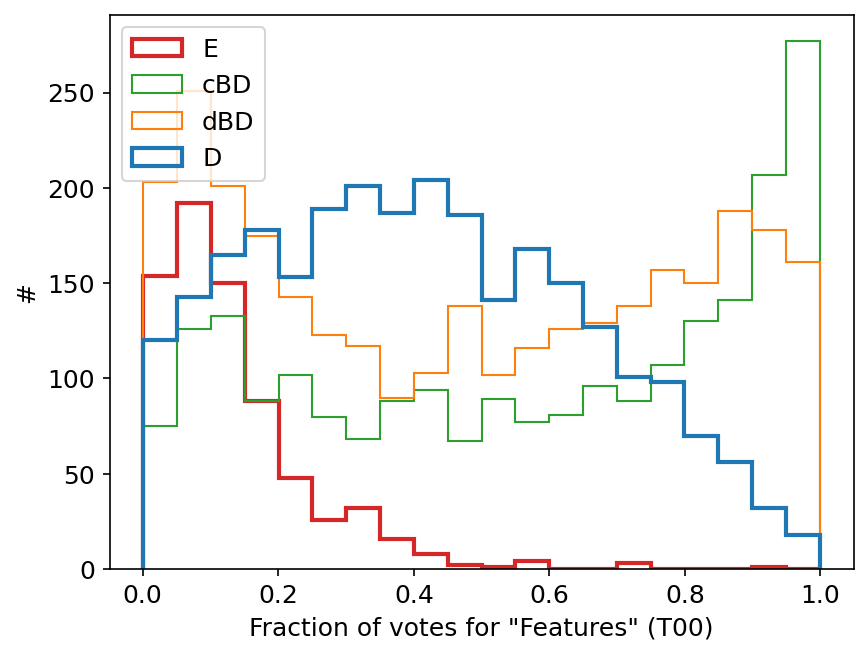}
    \caption{The fraction of voting for T00 question ``smooth or featured" for the visual classifications presented in \cite{Driver22}: Elliptical, compact-bulge (cBD), diffuse disk (dBD) and disk-dominated (D) galaxies. Ellipticals have the lowest voting fractions, followed by disk-dominated and both cBD and dBD have a high voting fraction for ``features''. }
    \label{f:viz:T00}
\end{figure}

\begin{figure}[htbp]
    \centering
    \includegraphics[width=\linewidth]{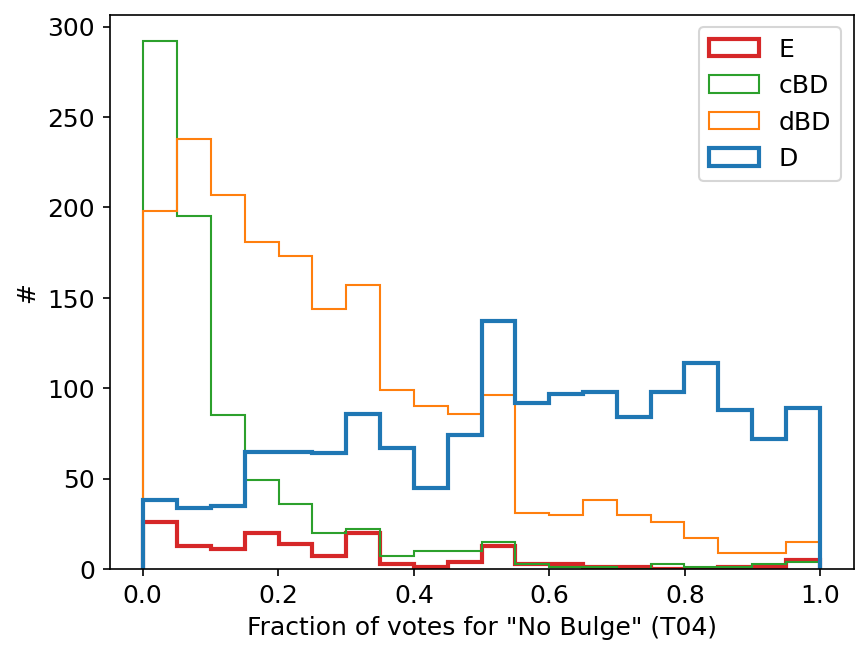}
    \caption{The ``no bulge'' fraction of voting for T04 question ``How prominent is the bulge?" for the visual classifications presented in \cite{Driver22}: Elliptical, compact-bulge (cBD), diffuse disk (dBD) and disk-dominated (D) galaxies. This is the other question that can be directly compared to the \cite{Driver22} classifications as these focus on the prominence of the bulge. Highest fraction is the ``pure disk'' (D), followed by the diffuse and concentrated bulge classes. Ellipticals are rarely in this question. }
    \label{f:viz:T04}
\end{figure}

There is a larger sample of overlap between the visual classifications by \cite{Driver22} and the KiDS-GZ catalogue. The classifications by \cite{Driver22} focus on the prominence of the bulge with respect to the galaxy as a whole. Both T00 and T04 are therefore good comparison questions (Table \ref{t:questions}, Figure \ref{f:flowchart}).

Figure \ref{f:viz:T00} shows the distribution of voting for the first question ``smooth or featured?''. Ellipticals have the lowest voting fractions, followed by disk-dominated classes (cBD, dBD and D). Both cBD and dBD have a high voting fraction for ``features'', higher than disk alone (D). 

The voting for T04 ``bulge prominence?'' in Figure \ref{f:viz:T04} for the answer ``no bulge''. The highest fraction is the ``pure disk'' (D), which is consistent with the KiDS-GZ vote. The diffuse bulge (dBD) is next, followed by the concentrated bulge (cBD).  Ellipticals are a small fraction of the galaxies in this question as most have been filtered out by T00.

Overall the expert classifications and the KiDS-GZ classifications agree well. Figures \ref{f:viz:T00} and \ref{f:viz:T04} can serve as a possible translation between Galaxy Zoo voting and expert classes (e.g. $0.2 < f_{T00} < 0.7$ and $f_{T04} > 0.5$ would select a fairly clean disk-only sample from the Galaxy Zoo voting.

\begin{figure}[htbp]
    \centering
    \includegraphics[width=\textwidth]{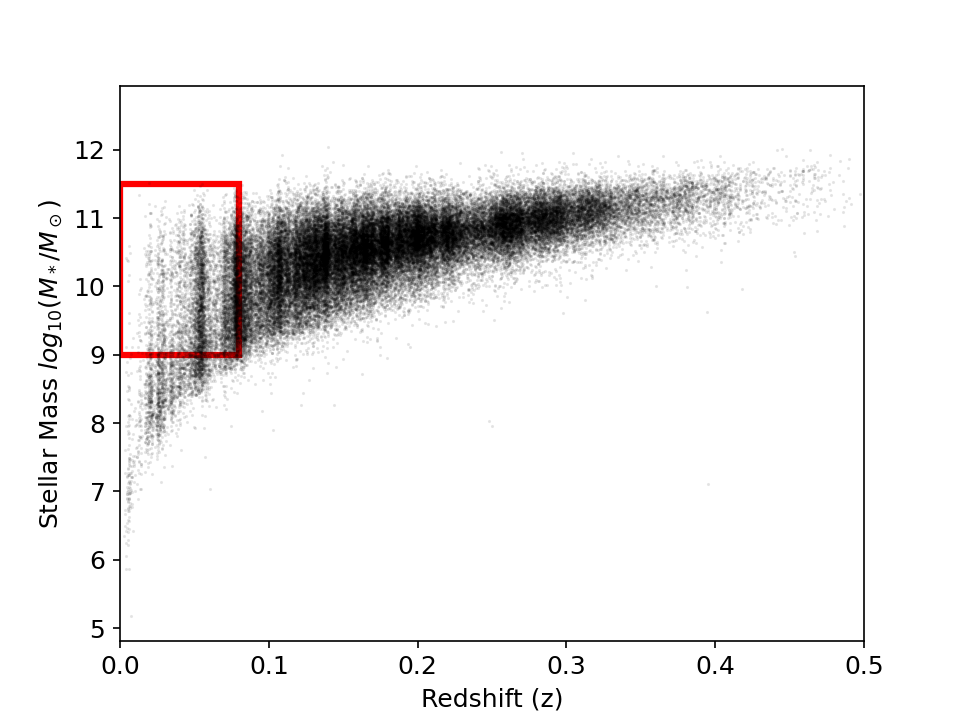}
    \caption{The redshift vs stellar mass as measured by {\sc magphys} for the GAMA galaxies with DESI-GZ \rev{{\sc zoobot} predictions} classifications. The red delineated area is the selection used in \cite{Porter-Temple22}. We select the same redshift range and stellar mass range for comparison but use the DESI-GZ \rev{{\sc zoobot} predictions} voting for the classification on number of spiral arms.}
    \label{f:PT:z-Mstar}
\end{figure}

\begin{figure}[htbp]
    \centering
    \includegraphics[width=\textwidth]{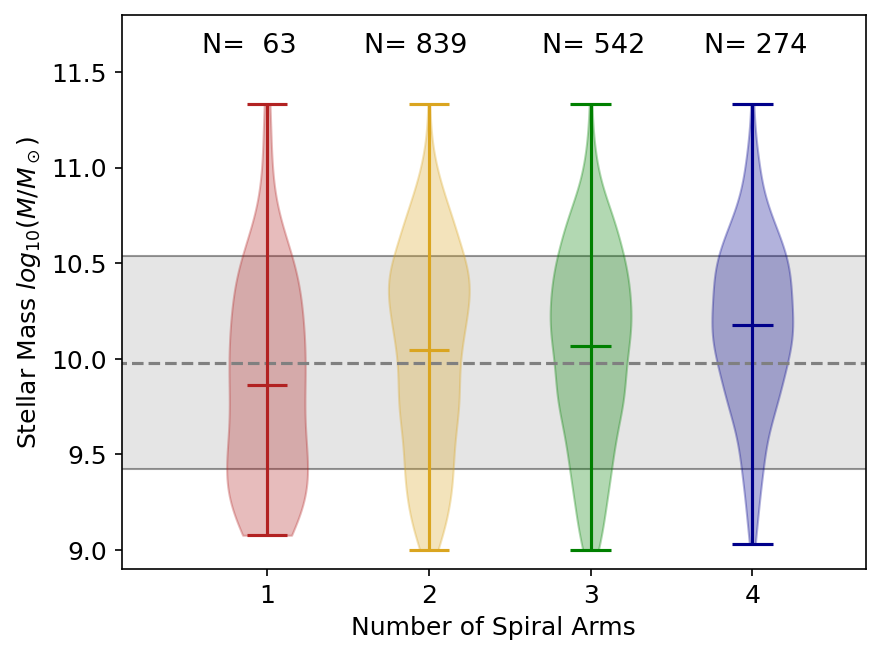}
    \caption{The distribution of stellar mass as measured by {\sc magphys} vs the number of spiral arms for the GAMA galaxies with DESI-GZ \rev{{\sc zoobot} predictions} classifications. }
    \label{f:PT:Mstar}
\end{figure}

\begin{figure}[htbp]
    \centering
    \includegraphics[width=\textwidth]{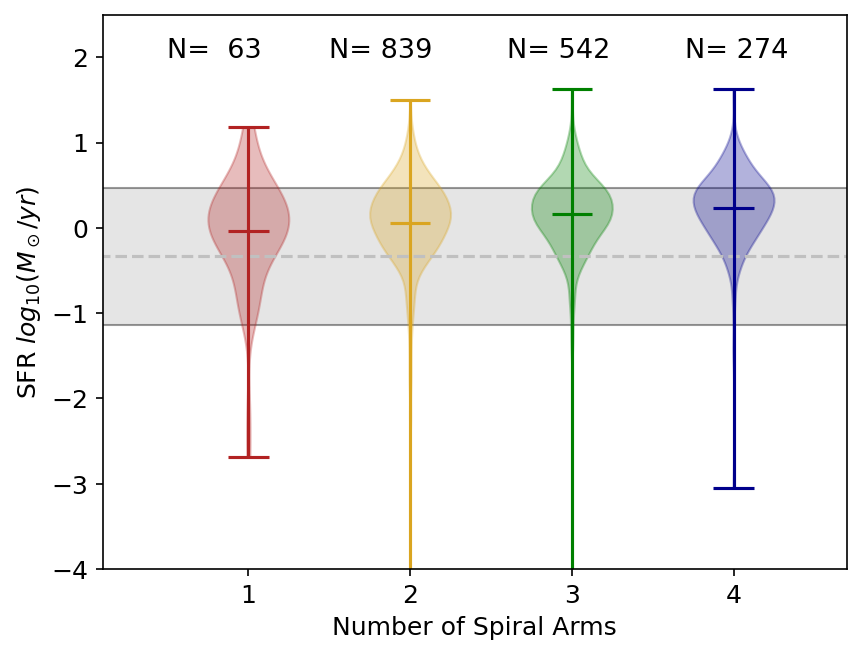}
    \caption{The distribution of star-formation rate as measured by {\sc magphys} vs the number of spiral arms for the GAMA galaxies with DESI-GZ \rev{{\sc zoobot} predictions} classifications.}
    \label{f:PT:SFR}
\end{figure}

\begin{figure}[htbp]
    \centering
    \includegraphics[width=\textwidth]{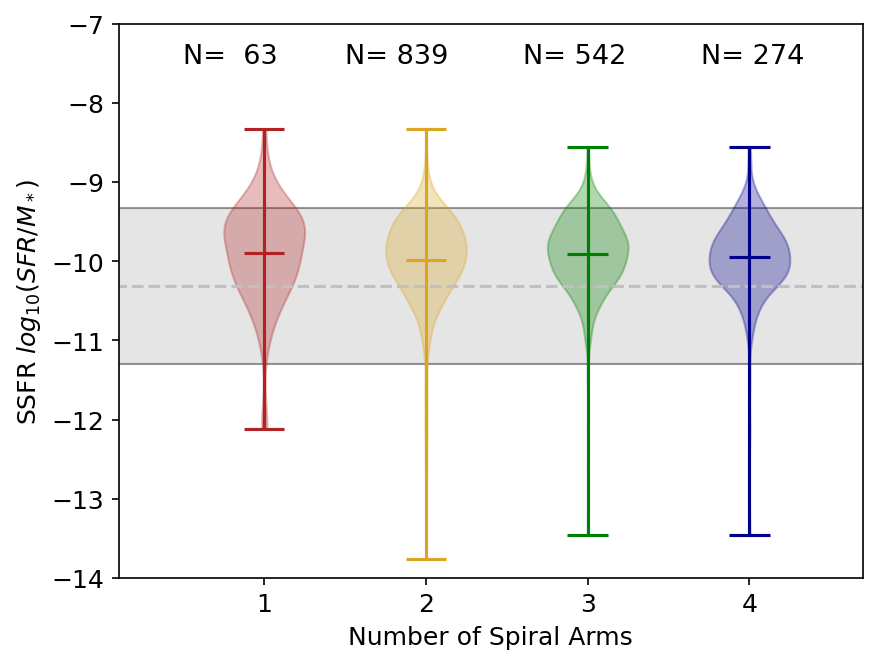}
    \caption{The distribution of the specific star-formation rate as measured by {\sc magphys} vs the number of spiral arms for the GAMA galaxies with DESI-GZ \rev{{\sc zoobot} predictions} classifications.}
    \label{f:PT:sSFR}
\end{figure}

\begin{table*}[htbp]
    \centering
    \begin{tabular}{l l l l l l l}
m & Stellar Mass  & P-T & SFR  & P-T          & sSFR & P-T \\
\hline
\hline
1 & 0.15 (0.1203)   & 0.138 (0.038) & 0.21 (0.0063) & 0.099 (0.256) & 0.23 (0.0023) & 0.234 (0.000)\\
2 & 0.08 (0.0011)   & 0.034 (0.203) & 0.25 (0.0000) & 0.110 (0.000) & 0.17 (0.0000) & 0.072 (0.000)\\
3 & 0.11 (0.0000)   & 0.152 (0.001) & 0.34 (0.0000) & 0.281 (0.000) & 0.21 (0.0000) & 0.187 (0.000)\\
4 & 0.22 (0.0000)   & 0.252 (0.081) & 0.40 (0.0000) & 0.291 (0.028) & 0.22 (0.0000) & 0.261 (0.064)\\
\hline
    \end{tabular}
    \caption{The Kolmogorov-Smirnov test and associated p-value for the number of arms (m). The comparison is between the population with m spiral arms and the full population. Stellar mass, specific star-formation and star-formation between each spiral arm number and the total population in Figure \ref{f:PT:z-Mstar}. Differences between the populations are similar in size and significance than they were in \cite{Porter-Temple22} using the KiDS-GZ classifications. Their K-S and p-values (their Table 1) are reproduced next to each column.}
    \label{t:KS}
\end{table*}

\section{Comparison to Porter-Temple+ (2022)}
\label{s:PTcomp}

Using the KiDS-GZ, \cite{Porter-Temple22} examined the dependence of stellar mass, star-formation rate, and specific star-formation rate with the number of spiral arms. They adopted a conservative approach in the identification of spiral arms by limiting the redshift to $z<0.08$, adopting a lower limit of $log(M_*/M_\odot) = 9$, and setting a relatively high threshold for a galaxy to be classified with one, two, three, four, or five and more spiral arms ($f>0.5$). 

Their selection criterion is shown as the red line box in Figure \ref{f:PT:z-Mstar}. The KiDS-GZ classifications were limited intentionally to z=0.15 but we can see from the GAMA galaxies with DESI-GZ \rev{{\sc zoobot} predictions} classifications in Figure \ref{f:PT:z-Mstar} that this limit is not enforced for DESI-GZ \rev{{\sc zoobot} predictions}. 
This is not the reason there is a much higher fraction in voting for smooth galaxies in DESI-GZ \rev{{\sc zoobot} predictions} (Figure \ref{f:T00}) because that sample is limited to z=0.15 by the crossmatch with KiDS-GZ. 
The images on which DESI-GZ \rev{{\sc zoobot} predictions} voting are based are shallower (Table \ref{t:depths}) and thus more prone to miss lower surface brightness features (spiral arms, tidal arms etc).


Voting in favor of T06 options other than 2-arms show slightly lower fractions for the same galaxies compared to the KiDS-GZ (Figure \ref{f:T06}, reflected in lower rankings as well). Therefore, we adopt slightly less stringent criteria to classify a galaxy with a certain number of spiral arms: we require that the $f_{disk}(T00) > 0.3$ and $f_{n-arm}(T06) > 0.2$ with $n$ the number of spiral arms. We also require the redshift to be $z<0.08$ as the DESI imaging is not higher resolution than KiDS and a minimum stellar mass of $log_{10}(M_*/M_\odot) > 9$. The lower voting fraction than \cite{Porter-Temple22} for a choice of $n$ arms is needed because otherwise the statistics for any number other than n=2 would be too low for a comparison. We note that 5+ still suffered from too low numbers to be included in the plot. 
 


Figure \ref{f:PT:Mstar} shows the distribution of stellar masses for n=1, 2, 3, or 4 spiral arms. The n=5+ category did not get enough votes for a statistically significant result. We see a similar rise in stellar mass with the number of spiral arms as \cite{Porter-Temple22}, compare to their Figure 4. 

Figure \ref{f:PT:SFR} shows the star-formation rate of galaxies with n=1, 2, 3, or 4 arms. Similar to \cite{Porter-Temple22}, their Figure 6, we see a rise in the star-formation with number of spiral arms, similar to the increase with mass. 

Figure \ref{f:PT:sSFR} shows the specific star-formation rate ($SFR/M_*$) of galaxies with n=1, 2, 3, or 4 arms. Similar to \cite{Porter-Temple22}, we see a flat or slight decline in the specific star-formation with number of spiral arms. A very similar, subtle decline in sSFR with the number of arms was observed by \cite{Porter-Temple22} in their Figure 8.

The comparison in stellar mass, star-formation, and specific star-formation can be done by comparing the distribution of values of galaxies with a certain number of arms (m) to the population as a whole. The similarity can be tested with the Kolmogorov-Smirnov test (K-S), which measures the greatest fractional difference in the cumulative distribution; 0 means no difference, and 1 means completely different distributions. The K-S test values are listed in Table \ref{t:KS} with the p-value in brackets. The differences in distributions are not very large, similar to what \cite{Porter-Temple22} found, and show the same trends.
The two-armed spiral, being the most numerous, will resemble the population at large the most, with the lowest K-S value. The trend is higher K-S values away from 2-arms. These are all the same trends observed by \cite{Porter-Temple22}. We conclude that with accurate inferred parameters --stellar mass, star-formation, and specific star-formation rates-- one can reproduce the experiment from \cite{Porter-Temple22} accurately.


\begin{figure}
    \centering
    \includegraphics[width=\textwidth]{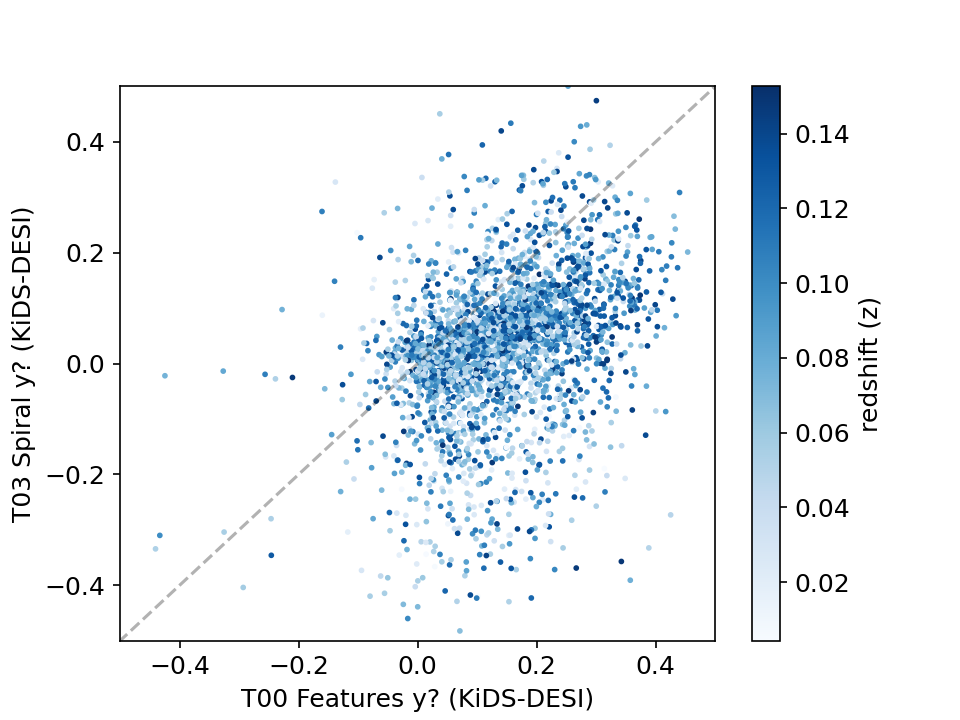}
    \caption{The difference in voting fraction between KiDS and DESI-based Galaxy Zoo on question T00 ``smooth or features?'' on the x-axis and T03 ``Spiral Structure?'' on the y-axis. The KiDS voting favors features over the DESI \rev{{\sc zoobot} predictions} but spiral structure is identified with similar frequency. T00 difference is not distance dependent and must be an inherent difference between DESI \rev{{\sc zoobot} predictions} and KiDS voting. }
    \label{f:T00:T03:comparison}
\end{figure}

\section{Discussion}
\label{s:discussion}

The two Galaxy Zoo iterations agree reasonably well with each other despite  different approaches to the imaging data that went into them and only slight differences in the classification questions. These were the same galaxies and observed in the same filters ($gri$), but on different telescopes, under different seeing conditions, to different depths, with a different approach to the generation of an RGB colour image, \rev{based on different numbers of classifiers, and in the case of DESI, extrapolated by {\sc zoobot}.} 

The correlation between answers (Table \ref{t:questions:stats}) show very good (linear) agreement between the voting fractions for most of the features. This adds to the confidence that these features are present in these galaxies, separately from the origin of the voting.
There is excellent agreement between KiDS-GZ and DES-GZ \rev{{\sc zoobot} predictions}. 


The question whether a volunteer sees features (T00) and whether they see spiral structure (T03) are somewhat overlapping; one would need to see features or disk structure to even see spiral arms. It is therefore perhaps illustrative to compare the difference in voting fractions in DESI/\rev{{\sc zoobot}} and KiDS-based Galaxy Zoo. We do so in Figure \ref{f:T00:T03:comparison}. There is a clear shift of voting in the KiDS-based Galaxy Zoo towards galaxies with features in T00 but then, once features have been found, the voting is mostly balanced around a difference of 0 for the T03. In the DESI \rev{{\sc zoobot} predictions}, more galaxies are identified as ``smooth'' but once features are identified, the result is similar to the KiDS one. This is not distance dependent and it is most likely the result of the depth of DESI compared to KiDS, \rev{influencing the final {\sc zoobot} predictions}. 

There is a lot of scatter in the fractions of votes. For individual galaxies, there may be room for interpretation, a well-known effect even among expert classifiers \citep[cf discussion in][]{Nair10a,Nair10b}. But for statistical uses, either catalogue looks to agree well with one another. Apart from perhaps removing unity values in the voting fraction, not much more correction is needed in KiDS-GZ. 

\rev{We check the KiDS-GZ voting with expert visual classifications by \cite{Stone23} and \cite{Driver22}. Both agree in broad terms with the voting in the Galaxy Zoo catalogue. The distributions of voting fractions for T00 and T04 agree with the categories assigned by experts, e.g., pure disk galaxies have a high voting fraction for no bulge and early types have a low voting fraction for no features. The broad agreement is another validation of the utility of this Galaxy Zoo catalogue for future uses. }

It is heartening to see that previous results by \cite{Porter-Temple22} are recovered here. The DESI-GZ \rev{{\sc zoobot} predictions} has voting for higher redshift galaxies and it is based on shallower imaging data, thus consistency with previous results strengthens its use case. Similarly, the KiDS-GZ voting fractions were not calibrated or de-biased but the higher thresholds compensate for that. Results like those in Figures \ref{f:PT:Mstar}--\ref{f:PT:sSFR} are only possible when the accuracy on both the x-axis i.e. the certainty in morphological classification and the accuracy on the y-axis i.e. the inferred galaxy property is of equally good quality, thanks to the multiwavelength photometry \citep{Wright16}. The voting in \cite{Hart17} was high accuracy with an earlier iteration of Galaxy Zoo but the accuracy in their star-formation measure did not quite match that of their Galaxy Zoo classifications, smoothing out the relation between arms and star-formation rate. 
The combination of voting and SED accuracy allowed \cite{Porter-Temple22} to improve on the \cite{Hart17} result. 
For similar reasons, we caution against the use of Galaxy Zoo questions on morphological details ($<$kpc in size) for redshifts over z=0.1 for either KiDS or DESI, as these correspond to more than the 0\farcs7 spatial resolution.

\section{Conclusions}
\label{s:conclusions}

In this paper, we directly compared two different iterations of the Galaxy Zoo morphology classification based on two different imaging surveys in the three GAMA equatorial fields. The images the classifications are based on, differ in depth, construction of RGB image, resolution, and target redshift range. \rev{The DESI-GZ catalogue is the result of {\sc zoobot} predictions based on all of DESI-LS trained classifications. }

We found that for individual galaxies, the voting fractions can often be quite different (several tens of percent; see Figures \ref{f:T00}--\ref{f:T09}). However, by and large the voting between both iterations agrees, especially for the populations at large. 

Reproducing the results from \citep{Porter-Temple22}, we find the same trends as they did using the DESI-GZ catalogue. With similar constraints on redshift, the DESI-GZ catalogue is suitable for similar work on morphological details. 

We note that the DESI-GZ \rev{{\sc zoobot} predictions} has a higher fraction of ``smooth'' classifications for galaxies that have more ``disk or features'' in T00. This is likely a combination of distance and depth of DESI imaging, hiding lower surface brightness features such as spiral arms and the disks of galaxies \rev{and an effect of weighting in the {\sc zoobot} classifications}.

\section{Acknowledgement}

The data in this paper are the result of the efforts of the Galaxy Zoo
volunteers, without whom none of this work would be possible. Their efforts are individually acknowledged at \url{http://authors.galaxyzoo.org}. 

This research made use of Astropy, a community-developed core Python package for Astronomy \citep{Astropy-Collaboration13,Astropy-Collaboration18}.


\newpage

\appendix

\section{Catalogue Descriptions}

Here we describe the GAMA KiDS and DESI based catalogues to accompany this paper. 

Tables \ref{t:t1} and \ref{t:t2} list the entries in the Galaxy Zoo KiDS classification catalog. These include a CATAID to identify the GAMA source and totals and fractions of voting on these objects. 

Table \ref{t:t3} is the full listing of entries in the DESI Galaxy Zoo catalogue as described in \cite{Walmsley23}. These the the GAMA CATAID and right ascention and declination used to match with the entries in the DESI catalogue. This catalogue contains only fractions of votes as these are predicted by zoobot. 

\begin{table*}[]
    \centering
    \begin{tabular}{l l l}
name            & unit          & description \\
\hline
\hline
region          & -             & The GAMA region \\
subject\_id     & -             & The identifier for subject. \\
survey\_id      & -             & The survey (GAMA) ID, identical to CATAID\\
zooniverse\_id  & -             & The Galaxy Zoo object identification \\
features\_total & $N_{T00}$     & Number of votes for T00 \\
edgeon\_total   & $N_{T01}$  & Number of votes for T01 \\
bar\_total      & $N_{T02}$     & Number of votes for T02 \\
spiral\_total   & $N_{T03}$   & Number of votes for T03 \\
bulge\_total    & $N_{T07}$   & Number of votes for T07 \\
spiralwinding\_total & $N_{T05}$   & Number of votes for T05 \\
spiralnumber\_total & $N_{T06}$   & Number of votes for T06 \\
bulgeshape\_total & $N_{T07}$ & Number of votes for T07 \\
round\_total    & $N_{T08}$   & Number of votes for T08 \\
mergers\_total  & $N_{T09}$   & Number of votes for T09 \\
oddtype\_total  & $N_{T10}$   & Number of votes for T10 \\
discuss\_total  & $N_{T11}$   & Number of votes for T11 \\
odd\_total      & $N_{votes}$   & Number of votes for T00 \\
\hline
    \end{tabular}
    \caption{The KiDS Galaxy Zoo catalogue entries. Total vote numbers for each question.}
    \label{t:t1}
\end{table*}

\begin{table*}[]
    \centering
    \begin{tabular}{l l l}
name            & name (unit)          & description \\
\hline
\hline
CATAID   & -  & The CATAID to match to GAMA catalogues.\\
\hline
features\_smooth\_frac              & $f_{smooth}$      & Fraction of votes T00 for ``smooth'' (A0)\\
features\_features\_frac            & $f_{features}$    & Fraction of votes T00 for ``features'' (A1)\\
features\_star\_or\_artifact\_frac  & $f_{artifact}$    & Fraction of votes T00 for ``star/artifact'' (A2)\\
\hline
edgeon\_yes\_frac                   & $f_{edge-on}$     & Fraction of votes T01 for yes edge-on (A0).\\
edgeon\_no\_frac                    & $f_{not edge-on}$ & Fraction of votes T01 for not edge-on (A1).\\
\hline
bar\_bar\_frac                      & $f_{bar}$         & Fraction of votes T02 for yes barred (A0).\\
bar\_no\_bar\_frac                  & $f_{no bar}$      & Fraction of votes T02 for no bar (A1).\\
\hline
spiral\_spiral\_frac                & $f_{spiral arms}$   & Fraction of votes T03 for yes spiral arms (A0).\\
spiral\_no\_spiral\_frac            & $f_{no spiral arms}$   & Fraction of votes T03 for no spiral arms (A1).\\
\hline
bulge\_no\_bulge\_frac              & $f_{no bulge}$   & Fraction of votes T04 for ``no bulge'' (A0).\\
bulge\_obvious\_frac                & $f_{obvious}$   & Fraction of votes T04 for ``obvious bulge'' (A1).\\
bulge\_dominant\_frac               & $f_{dominant}$   & Fraction of votes T04 for ``dominant bulge'' (A2).\\
\hline
spiralwinding\_tight\_frac          & $f_{tight}$   & Fraction of votes T05 for ``Tight spiral pattern'' (A0).\\
spiralwinding\_medium\_frac         & $f_{medium}$   & Fraction of votes T05 for ``Medium spiral pattern'' (A1).\\
spiralwinding\_loose\_frac          & $f_{loose}$   & Fraction of votes T05 for ``Loose spiral pattern'' (A2).\\
\hline
spiralnumber\_1\_frac               & $f_{1 arm}$   & Fraction of votes T06 for a single spiral arm (A0).\\
spiralnumber\_2\_frac               & $f_{2 arm}$   & Fraction of votes T06 for two spiral arms (A1).\\
spiralnumber\_3\_frac               & $f_{3 arm}$   & Fraction of votes T06 for three spiral arms (A2).\\
spiralnumber\_4\_frac               & $f_{4 arm}$   & Fraction of votes T06 for four spiral arms (A3).\\
spiralnumber\_more\_than\_4\_frac   & $f_{5+ arm}$   & Fraction of votes T06 for more than 4 spiral arms (A4).\\
\hline
bulgeshape\_rounded\_frac & $f_{round}$   & Fraction of votes T07 for a rounded bulge (A0).\\
bulgeshape\_boxy\_frac & $f_{boxy}$   & Fraction of votes T07 for a boxy bulge (A1).\\
bulgeshape\_no\_bulge\_frac & $f_{no bulge}$   & Fraction of votes T07 for no bulge (A2).\\
\hline                 
round\_completely\_round\_frac & $f_{round}$   & Fraction of votes T08 for round shape (A0).\\
round\_in\_between\_frac & $f_{in-between}$   & Fraction of votes T08 for in-between shape (A1).\\
round\_cigar\_shaped\_frac & $f_{cigar}$   & Fraction of votes T08 for cigar shape (A2).\\
\hline                 
mergers\_merging\_frac & $f_{merging}$   & Fraction of votes T09 for merging galaxies.\\
mergers\_tidal\_debris\_frac & $f_{debris}$   & Fraction of votes T09 for tidal debris or tails.\\
mergers\_both\_frac & $f_{both}$   & Fraction of votes T09 for both merging and tidal debris.\\
mergers\_neither\_frac & $f_{neither}$   & Fraction of votes T09 for neither merging or tidal debris.\\
\hline                 
oddtype\_none\_frac & $f_{no odd}$           & Fraction of votes T10 for no odd features (X0).\\
oddtype\_ring\_frac & $f_{ring}$           & Fraction of votes T10 for a ring feature (X1).\\
oddtype\_lens\_or\_arc\_frac & $f_{arc}$  & Fraction of votes T10 for lens/arc feature (X2).\\
oddtype\_irregular\_frac & $f_{irregular}$   & Fraction of votes T10 for irregular appearance (X3).\\
oddtype\_other\_frac & $f_{other}$   & Fraction of votes T10 for ``other'' features (X4).\\
oddtype\_dust\_lane\_frac & $f_{dust lane}$   & Fraction of votes T10 for a dust lane feature (X5).\\
oddtype\_overlapping\_frac & $f_{overlap}$   & Fraction of votes T10 for an overlapping pair (X6).\\
\hline                 
discuss\_yes\_frac & $f_{votes}$   & Fraction of votes T11 to flag for discussion.\\
discuss\_no\_frac & $f_{votes}$   & Fraction of votes T11 to not flag for discussion.\\
\hline
    \end{tabular}
    \caption{The KiDS Galaxy Zoo catalogue entries. Vote fraction for each question.}
    \label{t:t2}
\end{table*}


\begin{table*}[]
    \centering
    \begin{tabular}{l l l}
name            & name (unit)          & description \\
\hline
\hline
CATAID                                      & -             & GAMA catalog ID.\\
GAMA\_RA                                     & (degree)      & GAMA catalog right ascension.\\
GAMA\_DEC                                    & (degree)      & GAMA catalog declination.\\
dr8\_id                                      & -             & DESI DR8 ID.\\
ra                                          & (degree)      & DESI catalog right ascension.\\
dec                                         & (degree)      & DESI catalog declination.\\
brickid                                     & -             & DESI brick number \\
objid                                       & -             & DESI Object ID \\
hdf5\_loc                                    & -             & HDF5 location \\
\hline
smooth-or-featured\_smooth\_fraction          & $f_{smooth}$        & T00 fraction of votes for ``smooth'' (A0)\\
smooth-or-featured\_featured-or-disk\_fraction & $f_{featured}$     & T00 fraction of votes for ``featured'' (A1)\\
smooth-or-featured\_artifact\_fraction        & $f_{artifact}$      & T00 fraction of votes for ``artifact or star'' (A2)\\
\hline
disk-edge-on\_yes\_fraction                   & $f_{edgeon}$        & T01 fraction of votes for ``edgeon'' (A0) \\
disk-edge-on\_no\_fraction                    & $f_{edgeon}$        & T01 fraction of votes for ``not edgeon'' (A1) \\
\hline
has-spiral-arms\_yes\_fraction                & $f_{spiral}$        & T03 fraction of votes for ``spiral'' (A0) \\
has-spiral-arms\_no\_fraction                 & $f_{no-spiral}$     & T03 fraction of votes for ``no spiral'' (A1) \\
\hline
bar\_strong\_fraction                         & $f_{strongbar}$     & T03 fraction of votes for ``strong bar'' (A0) \\
bar\_weak\_fraction                           & $f_{weakbar}$       & T03 fraction of votes for ``weak bar'' (A1) \\
bar\_no\_fraction                             & $f_{nobar}$         & T03 fraction of votes for ``no bar'' (A2) \\
\hline
bulge-size\_dominant\_fraction                & $f_{dominant-bulge}$ & T04 fraction of votes for ``dominant bulge'' (A0) \\
bulge-size\_large\_fraction                   & $f_{large-bulge}$   & T04 fraction of votes for ``large bulge'' (A1) \\
bulge-size\_moderate\_fraction                & $f_{moderate-bulge}$ & T04 fraction of votes for ``moderate bulge'' (A2) \\
bulge-size\_small\_fraction                   & $f_{small-bulge}$   & T04 fraction of votes for ``small bulge'' (A3) \\
bulge-size\_none\_fraction                    & $f_{no-bulge}$      & T04 fraction of votes for ``no bulge'' (A4) \\
\hline
spiral-winding\_tight\_fraction               & $f_{tight}$         & T05 fraction of votes for ``tightly'' wound spiral arms (A0) \\
spiral-winding\_medium\_fraction              & $f_{medium}$        & T05 fraction of votes for ``medium'' wound spiral arms (A1) \\
spiral-winding\_loose\_fraction               & $f_{loose}$         & T05 fraction of votes for ``loosely'' wound spiral arms (A2) \\
\hline
spiral-arm-count\_1\_fraction                 & $f_{1-arm}$         & T06 fraction of votes for 1 spiral arm (A0) \\
spiral-arm-count\_2\_fraction                 & $f_{2-arm}$         & T06 fraction of votes for 2 spiral arms (A1) \\ 
spiral-arm-count\_3\_fraction                 & $f_{3-arm}$         & T06 fraction of votes for 3 spiral arms (A2) \\ 
spiral-arm-count\_4\_fraction                 & $f_{4-arm}$         & T06 fraction of votes for 4 spiral arms (A3) \\ 
spiral-arm-count\_more-than-4\_fraction       & $f_{5+ arm}$        & T06 fraction of votes for 5+ spiral arms (A4) \\ 
spiral-arm-count\_cant-tell\_fraction         & $f_{?-arm}$         & T06 fraction of votes for unclear number of spiral arms (A5)\\
\hline
edge-on-bulge\_boxy\_fraction                 & $f_{round}$         & T07 fraction of votes for ``rounded'' bulge (A0) \\
edge-on-bulge\_none\_fraction                 & $f_{round}$         & T07 fraction of votes for ``boxy'' bulge (A1) \\
edge-on-bulge\_rounded\_fraction              & $f_{round}$         & T07 fraction of votes for ``round'' galaxy (A2) \\
\hline
how-rounded\_round\_fraction                  & $f_{round}$         & T08 fraction of votes for ``round'' galaxy (A0) \\
how-rounded\_in-between\_fraction             & $f_{in-between}$    & T08 fraction of votes for ``in-between'' (A1) \\
how-rounded\_cigar-shaped\_fraction            & $f_{cigar}$         & T08 fraction of votes for ``cigar'' (A2) \\
\hline
merging\_none\_fraction                       & $f_{merging}$       & T09 fraction of votes for ``merging'' (A0) \\
merging\_minor-disturbance\_fraction          & $f_{tidal}$       & T09 fraction of votes for ``tidal debris'' (A1) \\
merging\_major-disturbance\_fraction          & $f_{both}$       & T09 fraction of votes for ``both'' merging and tidal debris (A2) \\
merging\_merger\_fraction                     & $f_{neither}$       & T09 fraction of votes for ``neither'' merging or tidal debris (A3) \\
\hline
    \end{tabular}
    \caption{The DESI Galaxy Zoo catalogue entries. CATAID and basic information from GAMA target catalogue and the vote fraction for each question.}
    \label{t:t3}
\end{table*}

\end{document}